\documentclass[twocolumn]{autart}    % Enable this line and disable the
\usepackage{graphicx}          % Include this line if your
\usepackage{amsmath,amsfonts,amssymb,mathrsfs}
\usepackage{url}
\usepackage{color}
\usepackage{multirow}
\usepackage{algorithm}
\usepackage{algpseudocode}
\renewcommand\qed{$\blacksquare$}

\newtheorem{theorem}{Theorem}
\newtheorem{definition}{Definition}
\newtheorem{proposition}{Proposition}
\newtheorem{lemma}{Lemma}
\newtheorem{corollary}{Corollary}
\newtheorem{remark}{Remark}
\newtheorem{example}{Example}
\newtheorem{counterexample}{Counterexample}
\newtheorem{property}{Property}

\begin{document}

\begin{frontmatter}

\title{Probabilistic Model Validation for Uncertain Nonlinear Systems\thanksref{footnoteinfo}} % Title, preferably not more

\thanks[footnoteinfo]{The material in this paper was partially presented in IEEE Conference on Decision and Control (CDC) 2011 \cite{HalderBhattacharyaCDC2011} and 2012 \cite{HalderBhattacharyaCDC2012}. Corresponding author: A.~Halder. Tel. +979-583-6070, Fax +979-845-6051.}

\author[]{Abhishek Halder}\ead{ahalder@tamu.edu},    % Add the
\author[]{Raktim Bhattacharya}\ead{raktim@tamu.edu}               % e-mail address

\address{Department of Aerospace Engineering, Texas A\&M University, College Station, TX 77843-3141, United States}  % Please supply

                             % keyword list or with the
                                          % help of the Automatica
                                          % keyword wizard

\begin{abstract}                          % Abstract of not more than 200 words.
This paper presents a probabilistic model validation methodology for nonlinear systems in time-domain. The proposed formulation is simple, intuitive, and accounts both deterministic and stochastic nonlinear systems with parametric and nonparametric uncertainties. Instead of hard invalidation methods available in the literature, a relaxed notion of validation in probability is introduced. To guarantee provably correct inference, algorithm for constructing probabilistically robust validation certificate is given along with computational complexities. Several examples are worked out to illustrate its use.
\end{abstract}

\end{frontmatter}

\section{Introduction}
A model serves as a mathematical abstraction of the physical system, providing a framework for system analysis and controller synthesis. Since such mathematical representations are based on assumptions specific to the process being modeled, it's important to quantify the reliability to which the model is consistent with the physical observations. Model quality assessment is imperative for applications where the model needs to be used for prediction (e.g. weather forecasting, stock market) or safety-critical control design (e.g. aerospace, nuclear, systems biology) purposes.

Here it is important to realize that a model can only be validated against experimental observations, not against another model. Thus a \emph{model validation problem} can be stated as: \emph{given a candidate model and experimentally observed measurements of the physical system, how well does the model replicate the experimental measurements?} It has been argued in the literature \cite{Popper1963,SmithDoyle1992,Poolla1994,Prajna2006} that the term `model validation' is a misnomer since it would take infinite  number of experimental observations to do so. Hence the term `model invalidation' or `falsification' \cite{BrugarolasSafonov2002} is preferred. In this paper, instead of hard invalidation, we will consider the validation/invalidation problem in a probabilistically relaxed sense.

\subsection{Related literature}
Broadly speaking, there have been three distinct frameworks in which the model validation problem has been attempted till now. \textbf{One} is a discrete formulation in \emph{temporal logic framework} \cite{BaierKatoen2008} which has been extended to account probabilistic models \cite{BaierKatoen2008,CiesinskiGrosser2004}. \textbf{Second} is the $\mathcal{H}_{\infty}$ \emph{control framework} where time-domain \cite{Poolla1994,SmithDullerud1996,ChenWang1996}, frequency domain \cite{SmithDoyle1992,WahlbergLjung1992} and mixed domain \cite{Xu1999} model validation methods have been studied assuming structured norm-bounded uncertainty in linear dynamics setting. The \textbf{third} framework involves deductive inference based on barrier certificates \cite{Prajna2006} which was shown to encompass a large class of nonlinear models including differential-algebraic equations \cite{Campbell1980}, dynamic uncertainties described by integral quadratic constraints \cite{MegretskiRantzer1997}, stochastic \cite{Oksendal2003} and hybrid dynamics \cite{vanderSchaftSchumacher1999}.

In statistical setting, model validation has been addressed from system identification perspective \cite{LjungGuo1997,Ljung1999} where the main theme is to validate an identified nominal model through correlation analysis of the residuals. A polynomial chaos framework has also been proposed \cite{Ghanem2008} for model validation. Gevers \emph{et. al.} \cite{Gevers2003} have connected the robust control framework with prediction error based identification for frequency-domain validation of linear systems. In another vein, using Bayesian conditioning, Lee and Poolla \cite{LeePoolla1996} showed that for \emph{parametric} uncertainty models, the statistical validation problem may be reduced to the computation of relative weighted volumes of convex sets. However, for \emph{nonparametric} models: ``the situation is significantly more complicated" \cite{LeePoolla1996} and to the best of our knowledge, has not been addressed in the literature. Recently, in the spirit of weak stochastic realization problem \cite{vanSchuppen1989}, Ugrinovskii \cite{Ugrinovskii2009} investigated the conditions for which the output of a stochastic nonlinear system can be realized through perturbation of a nominal stochastic \emph{linear} system.

In practice, one often encounters the situation where a model is either proposed from physics-based reasoning or a reduced order model is derived for computational convenience. In either case, the model can be linear or nonlinear, continuous or discrete-time, and in general, it's not possible to make any a-priori assumption about the noise. Given the experimental data and such a candidate model for the physical process, our task is to answer: ``to what extent, the proposed model is valid?" In addition to quantify such degree of validation, one must also be able to demonstrate that the answer is \emph{provably correct} in the face of uncertainty. This brings forth the notion of \emph{probabilistically robust model validation}. In this paper, we will show how to construct such a \emph{robust validation certificate}, guaranteeing the performance of probabilistic model validation algorithm.

\subsection{Contributions of this paper}

With respect to the literature, the contributions of this paper are as follows.
\begin{enumerate}
\item Instead of interval-valued structured uncertainty (as in $\mathcal{H}_{\infty}$ control framework) or moment based uncertainty (as in parametric statistics framework), this paper deals with model validation in the sense of nonparametric statistics. Uncertainties in the model are quantified in terms of the probability density functions (PDFs) of the associated random variables. We argue that such a formulation offers several advantages. \emph{Firstly}, we show that model uncertainties in the parameters, initial states and input disturbance, can be propagated accurately by spatio-temporally evolving the joint state and output PDFs. Since experimental data usually come in the form of histograms, it's a more natural quantification of uncertainty than specifying sets \cite{Prajna2006} to which the trajectories are contained at each instant of time. However, if needed, such sets can be recovered from the supports of the instantaneous PDFs. \emph{Secondly}, as we'll see in Section 5, instead of simply invalidating a model, our methodology allows to estimate the probability that a proposed model is valid or invalid. This can help to decide which specific aspects of the model need further refinement. Hard invalidation methods don't cater such constructive information. \emph{Thirdly}, the framework can handle both discrete-time and continuous-time nonlinear models which need not be polynomial. Previous work like \cite{Prajna2006} dealt with semialgebraic nonlinearities and relied on sum of squares (SOS) decomposition \cite{Parrilo2000} for computational tractability. From an implementation point of view, the approach presented in this paper doesn't suffer from such conservatism.

\item Due to the uncertainties in initial conditions, parameters, and process noise, one needs to compare output ensembles instead of comparing individual output realizations. This requires a metric to quantify closeness between the experimental data and the model in the sense of distribution. We propose \emph{Wasserstein distance} to compare the output PDFs and argue why commonly used information-theoretic notions like \emph{Kullback-Leibler divergence} may not be appropriate for this purpose.

\item We show that the uncertainty propagation through continuous or discrete-time dynamics can be done via numerically efficient meshless algorithms, even when the model is high-dimensional and strongly nonlinear. Moreover, we outline how to compute the Wasserstein distance in such settings. Further, bringing together ideas from analysis of randomized algorithms, we give sample-complexity bounds for robust validation inference.
\end{enumerate}

The paper is organized as follows. In Section 2, we describe the problem setup. Then we expound on the three steps of our validation framework, viz. uncertainty propagation, distributional comparison and construction of validation certificates in Section 3, 4 and 5, respectively. We provide numerical examples in Section 6, to illustrate the ideas presented in this paper. The concept of worst-case initial uncertainty related to model discrimination, is addressed in Section 7. Section 8 presents some results for discrete-time linear Gaussian systems, followed by conclusions in Section 9.

\begin{figure*}[htb]
\begin{center}
\vspace*{-1in}
\hspace*{-0.7in}
\includegraphics[width=7.3in]{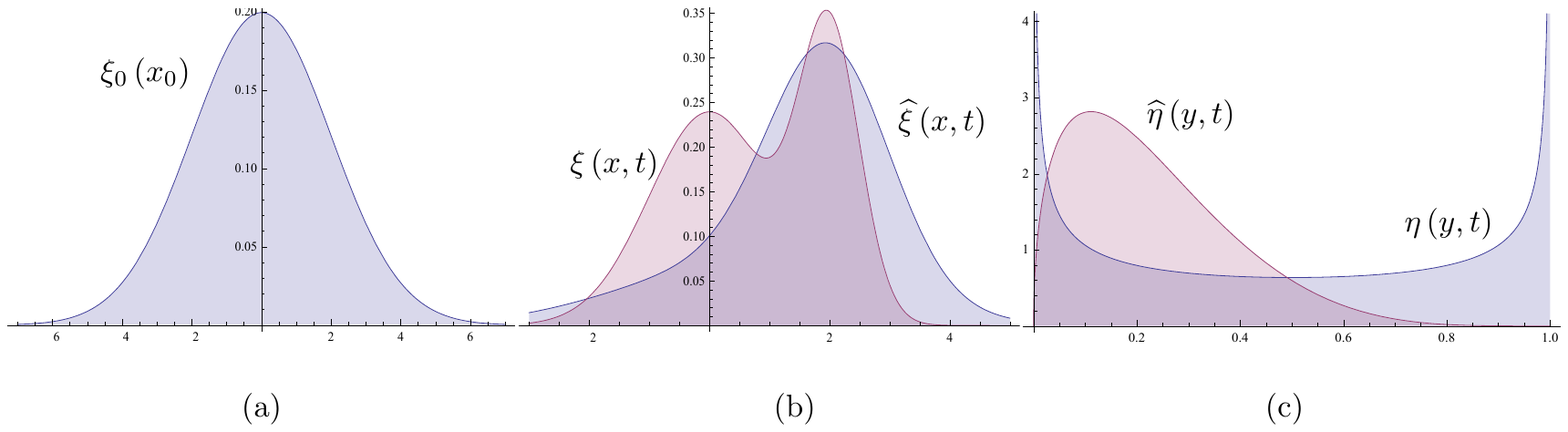}
\end{center}
\vspace*{-7.6in}
\caption{\small{The proposed model validation framework compares experimentally observed output PDF $\eta\left(y,t\right)$ with the model-predicted output PDF $\widehat{\eta}\left(\widehat{y},t\right)$, the comparison being made with respect to some suitable metric at each instant of measurement availability. The state dynamics evolves the initial joint PDF $\xi_{0}\left(x_{0}\right)$ (Fig. 1(a)) to instantaneous joint state PDFs $\xi\left(x,t\right)$ and $\widehat{\xi}\left(\widehat{x},t\right)$ (Fig. 1(b)). The associated output PDFs $\eta\left(y,t\right)$ and $\widehat{\eta}\left(y,t\right)$ may share the
same support ($\left[0,1\right]$ as shown in Fig. 1(c)), but have different shapes. Hence, instead of matching output supports, we propose
matching output PDFs at all times, for validating a model.}}%
\label{intuition1}%
\end{figure*}

\subsection*{Notation}
We use the superscript $^{\top}$ to denote matrix transpose, $\otimes$ to denote Kronecker product, and the symbol $\wedge$ to denote minimum of two real numbers. The notation $_{r}F_{s}\left(a_{1}, \hdots, a_{r}; b_{1}, \hdots, b_{s}; x\right)$ stands for generalized hypergeometric function. The symbols $\mathcal{N}\left(.,.\right)$, $\mathcal{U}\left(.\right)$, and $\mathcal{A}\left(.\right)$ are used for normal, uniform and arcsine distributions, respectively. We use the notation $\xi_{0}\left(.\right)$ to denote the joint PDF over initial states and parameters. $\xi\left(.,t\right)$ and $\widehat{\xi}\left(.,t\right)$ denote joint PDFs over instantaneous states and parameters, for the true and model dynamics, respectively. Similarly, $\eta\left(.,t\right)$ and $\widehat{\eta}\left(.,t\right)$, respectively denote joint PDFs over output spaces $y$ and $\widehat{y}$ at time $t$, for the true and model dynamics. The symbol $\widetilde{x}$ is used to denote the extended state vector obtained by augmenting the state ($x$) and parameter ($p$) vectors. We use $\chi$ to denote indicator function and \# to denote cardinality. Unless stated otherwise, $\delta\left(.\right)$ stands for Dirac delta. The symbol $I_{\ell}$ denotes the $\ell$-by-$\ell$ identity matrix, $\nabla_{x}$ denotes gradient operator with respect to vector $x$, $\text{vec}\left(\cdot\right)$ stands for the vectorization operator, and $\parallel \cdot \parallel_{F}$ denotes the Frobenius norm. $\text{tr}\left(\cdot\right)$ and $\text{det}\left(\cdot\right)$ stand for trace and determinant of a matrix. The abbreviations \emph{a.s.} and \emph{i.p.} refer to convergence in \emph{almost sure} and \emph{in probability} sense. The shorthand $\partial_{\alpha}$ means partial derivative with respect to variable $\alpha$, $\text{supp}\left(\cdot\right)$ denotes support of a function, and $\text{erf}(\cdot)$ stands for error function.

\section{Problem Setup}

\subsection{Intuitive idea}

The proposed framework is based on the evolution of densities in output space, instead of evolution of individual trajectories, as in the Lyapunov framework. Intuitively, characteristics of the input to output mapping is revealed by the growth or depletion of trajectory concentrations in the output-space. Growth in concentration, or increased density, defines regions in where the trajectories accumulate. This corresponds to regions with slow time scale dynamics or time invariance. Similarly, depletion of concentration in a set implies fast-scale dynamics or unstable manifold. We refer the readers to \cite{LasotaMackey1994} for an introduction to analysis of dynamical systems using trajectory densities. This idea of comparing dynamical systems based on density functions, have been presented before by Sun and Mehta \cite{SunMehta2010} in the context of filtering, and by Georgiou \cite{georgiou2007distances} in the context of matching power spectral densities.

\subsubsection{Proposed approach}
Given the experimental measurements of the physical system in the form of a time-varying distribution (such as histograms), we propose to compare the \emph{shape} or \emph{concentration profile} of this measured output density, with that predicted by the model. At every instant of time, if the model-predicted density matches with the experimental one ``\emph{reasonably well}" (to be made precise later in the paper), we conclude that the model is validated with high \emph{confidence} (to be computed for guaranteeing quality of inference).

\subsubsection{Why compare densities instead of trajectories}
The rationale behind comparing the distributional shapes for model validation comes from the fact that the presence of uncertainties mask the difference between individual output realizations. Uncertainties in initial conditions, parameters and noise result different realizations of the trajectory or integral curve of the dynamical system. Regions of high (low) concentration of trajectories correspond to regions of high (low) probability. Thus a model validation procedure should naturally aim to compare concentrations of the trajectories between the measurements and model-predictions, instead of comparing individual realizations of them, which would be meaningful only in the absence of uncertainties.

We would like to point out that in some applications, the measurement naturally arises in the form of a distribution. This includes (i) \textbf{process industry applications} like measurement made at the wet end of papermaking machines \cite{Wang1999,WangBakiKabore2001} that involves the fibre length and filler size distribution sensed via vision sensors, (ii) \textbf{Nuclear Magnetic Resonance (NMR) spectroscopy and Imaging (MRI) applications} where the measurement variable is magnetization distribution \cite{LiKhaneja2009}, (iii) \textbf{neuroscience applications} where the measurement variable is the distribution of frequency across a collection of neurons \cite{BrownMoehlisHolmes2004}, and (iv) \textbf{social systems} where the measurement variable could be an ensemble of crowd \cite{WadooKachroo2010} sensed via cameras or motion detectors. Notice that for (i) and (iii), distributional measurement is a design choice; for (ii) it is motivated by technological limitations of sensing individual magnetization states where the number of states are of the order of Avogadro number $6 \times 10^{23}$; and for (iv) individual measurement may raise privacy concerns.

\subsubsection{Why compare densities instead of moments or sets}
Density based model validation provides natural advantages over moment based or set containment methods for the following reasons. Moment based methods can be erroneous for nonlinear non-Gaussian systems, as two different trajectory densities may provide the same correlation information. This can be circumvented by including higher order moments, but it is not computationally tractable for high dimensional systems. Set containment arguments can also be erroneous as it is possible that at a given time, two systems have trajectory densities with identical supports but different concentrations (Fig. \ref{intuition1} (c)).

A proposed model is validated, if the ``distance" between its predicted density and the measured density, remains below a user-specified tolerance level, which need not be fixed over time. For example, take-off and landing are critical operational segments during the flight of a commercial aircraft, and it's unacceptable to have a controller that does not guarantee the robust performance for these critical time-segments with very high probability. This motivates the computation of probability of validation as part of the model validation oracle.

\begin{figure}[t]
\vspace*{-0.15in}
\begin{center}
 \includegraphics[scale=0.85]{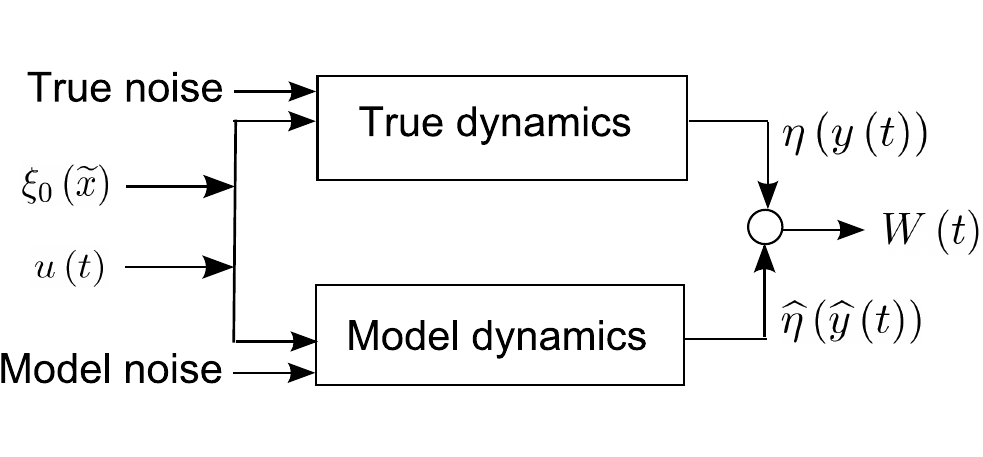}
 \end{center}
 \vspace*{-0.15in}
\caption{Block diagram for the proposed model validation formulation.}
\label{BlockDiagram}
\end{figure}

\subsection{Methodology}
In this section, we formalize the ideas presented above. Fig. \ref{intuition1} and \ref{BlockDiagram} show the outline of the model validation framework proposed here. In this formulation, the systems under comparison are excited with a \emph{known} input signal $u\left(t\right)$, and an initial PDF $\xi_{0}\left(\widetilde{x}_{0}\right)$, supported over the extended state space $\widetilde{x} := \{x, \; p\}^{\top}$, where the states $x \in \mathbb{R}^{n_{s}}$, and the parameters $p \in \mathbb{R}^{n_{p}}$. Given the PDF $\eta\left(y\left(t\right)\right)$ supported over the true output space $y \in \mathbb{R}^{n_{o}}$, and a candidate model, we compute and then compare the model predicted output PDF $\widehat{\eta}\left(\widehat{y}\left(t\right)\right)$, with $\eta\left(y\left(t\right)\right)$ at each instances of measurement availability $\{t_{j}\}_{j=1}^{\tau}$. Thus, one can think of three distinct steps of such a model validation framework. These are:
\begin{enumerate}
 \item evolving $\xi_{0}\left(\widetilde{x}_{0}\right)$ using the proposed model, to compute $\widehat{\eta}\left(\widehat{y}\left(t\right)\right)$,
 \item measuring an appropriate notion of distance, denoted as $W(t)$ in Fig. \ref{BlockDiagram}, between $\eta\left(y\left(t\right)\right)$ and $\widehat{\eta}\left(\widehat{y}\left(t\right)\right)$ at $\{t_{j}\}_{j=1}^{\tau}$,
 \item probabilistic quantification of provably correct inference in this framework and providing sample complexity bounds for the same.
\end{enumerate}
Now we will elicit each of these steps.

\section{Uncertainty Propagation}
\subsection{Continuous-time models}
\subsubsection{Uncertainty propagation for deterministic flow}
Consider the continuous-time nonlinear model with state dynamics given by the ODE $\dot{\widehat{x}} = \widehat{f}\left(\widehat{x}, \widehat{p}\right)$, where $\widehat{x}\left(t\right) \in \widehat{\mathcal{X}} \subseteq \mathbb{R}^{\widehat{n}_{s}}$ is the state vector, $\widehat{p}\in \widehat{\mathcal{P}} \subseteq \mathbb{R}^{\widehat{n}_{p}}$ is the parameter vector, the dynamics $\widehat{f}\left(.,\widehat{p}\right) : \widehat{\mathcal{X}} \mapsto \mathbb{R}^{\widehat{n}_{s}}$ $\forall \: \widehat{p} \in \widehat{\mathcal{P}}$, and is at least locally Lipschitz . It can be put in an extended state space form
\begin{eqnarray}
\dot{\widehat{\widetilde{x}}} = \widehat{\widetilde{f}}\left(\widehat{\widetilde{x}}\right), \; \widehat{\widetilde{x}} \in \widehat{\mathcal{X}} \times \widehat{\mathcal{P}} \subseteq \mathbb{R}^{\widehat{n}_{s}+ \widehat{n}_{p}}, \; \widehat{\widetilde{f}} = \begin{Bmatrix}\widehat{f}_{\widehat{n}_{s}\times 1} \\ \mathbf{0}_{\widehat{n}_{p}\times 1}\end{Bmatrix}.
\label{ExtendedStateSpace}
\end{eqnarray}
The output equation can be written as
\begin{eqnarray}
\widehat{y} = \widehat{h}\left(\widehat{\widetilde{x}}\right), \quad \widehat{h}: \widehat{\mathcal{X}} \times \widehat{\mathcal{P}} \mapsto \widehat{\mathcal{Y}},
\label{OutputDynamics}
\end{eqnarray}
where $\widehat{y}\left(t\right) \in \widehat{\mathcal{Y}} \subseteq \mathbb{R}^{n_{o}}$ is the output vector. If uncertainties in the initial conditions $\left(x_{0}:=x\left(0\right)\right)$ and parameters $\left(\widehat{p}\right)$ are specified by the initial joint PDF $\xi_{0}\left(\widetilde{x}\right)$, then the evolution of uncertainties subject to the dynamics (\ref{ExtendedStateSpace}), can be described by evolving the joint PDF $\widehat{\xi}\left(\widehat{\widetilde{x}}, t\right)$ over the extended state space. Such spatio-temporal evolution of $\widehat{\xi}\left(\widehat{\widetilde{x}}, t\right)$ is governed by the \emph{stochastic Liouville equation} (SLE) given by (Section 7.6 in \cite{LasotaMackey1994})
\begin{eqnarray}
\displaystyle\frac{\partial \widehat{\xi}}{\partial t} &=& \mathscr{L}_{\text{SLE}}\widehat{\xi} = D_{1} \widehat{\xi} = -\nabla . \left(\widehat{\xi} \widehat{f}\right) = -\displaystyle\sum_{i=1}^{\widehat{n}_{s}}\displaystyle\frac{\partial}{\partial \widehat{x}_{i}} \left(\widehat{\xi} \widehat{f}_{i}\right),
\label{Liouville}
\end{eqnarray}
which is a quasi-linear partial differential equation (PDE), first order in both space and time. Notice that, the spatial operator $\mathscr{L}_{\text{SLE}}$ is a drift operator $D_1$ that describes the \emph{advection} of the PDF in extended state space. The output PDF $\widehat{\eta}\left(\widehat{y}, t\right)$ can be computed from the state PDF as
\begin{eqnarray}
\widehat{\eta}\left(\widehat{y}, t\right) = \displaystyle\sum_{j=1}^{\nu} \:\displaystyle\frac{\widehat{\xi}\left(\widehat{\widetilde{x}}_{j}^{\star}\right)}
{|\text{det}\left(\mathcal{J}\left(\widehat{\widetilde{x}}_{j}^{\star}\right)\right)|},
\label{OutputPDF}
\end{eqnarray}
where $\widehat{\widetilde{x}}_{j}^{\star}$ is the $j$\textsuperscript{th} root of the inverse transformation of (\ref{OutputDynamics}) with $j = 1, 2, \hdots, \nu$, and $\mathcal{J}$ is the Jacobian of this inverse transformation.

\subsubsection{Uncertainty propagation for stochastic flow}
Consider the continuous-time nonlinear model with state dynamics given by the It$\hat{\text{o}}$ SDE
\begin{eqnarray}
d\widehat{\widetilde{x}} = \widehat{\widetilde{f}}\left(\widehat{\widetilde{x}}\right) \: dt + \widehat{g}\left(\widehat{\widetilde{x}}\right) \:d\beta,
\label{ItoSDE}
\end{eqnarray}
where $\beta\left(t\right) \in \mathbb{R}^{\omega}$ is the $\omega$-dimensional Wiener process at time $t$, and the noise coupling $\widehat{g} : \widehat{\mathcal{X}} \times \widehat{\mathcal{P}} \mapsto \mathbb{R}^{(\widehat{n}_{s} + \widehat{n}_{p}) \times \omega}$. For the Wiener process $\beta\left(t\right)$, at all times
\begin{eqnarray}
\mathbb{E}\left[d\beta_{i}\right] = 0, \; \mathbb{E}\left[d\beta_{i}d\beta_{j}\right] = Q_{ij} = \alpha_{i} \: \delta_{ij} \; \forall \: i, j = 1,\hdots,\omega,
\label{WienerProcess}
\end{eqnarray}
where $\mathbb{E}\left[ . \right]$ stands for the expectation operator and $\delta_{ij}$ is the Kronecker delta. Thus $Q \in \mathbb{R}^{\omega \times \omega}$ with $\alpha_{i} > 0 \; \forall \: i=1, 2, \hdots, \omega$, being the noise strength. The output map is still assumed to be given by (\ref{OutputDynamics}). In such a setting, the evolution of the state PDF $\widehat{\xi}\left(\widehat{\widetilde{x}}, t\right)$ subject to (\ref{ItoSDE}), is governed by the \emph{Fokker-Planck equation} (FPE), also known as \emph{forward Kolmogorov equation}
\begin{eqnarray}
&&\displaystyle\frac{\partial \widehat{\xi}}{\partial t} = \mathscr{L}_{\text{FPE}}\widehat{\xi} = \left(D_{1} \: + D_{2}\right) \widehat{\xi} \nonumber\\
&=&  -\displaystyle\sum_{i=1}^{\widehat{n}_{s}}\displaystyle\frac{\partial}{\partial \widehat{x}_{i}} \left(\widehat{\xi} \widehat{f}_{i}\right) + \displaystyle\sum_{i=1}^{\widehat{n}_{s}}\displaystyle\sum_{j=1}^{\widehat{n}_{s}}\displaystyle\frac{\partial^{2}}{\partial \widehat{x}_{i}\partial \widehat{x}_{j}} \left(\left(\widehat{g}Q\widehat{g}^{\top}\right)_{ij} \widehat{\xi}\right),
\label{FokkerPlanck}
\end{eqnarray}
which is a homogeneous parabolic PDE, second order in space and first order in time. In this case, the spatial operator $\mathscr{L}_{\text{FPE}}$ can be written as a sum of a \emph{drift operator} $\left(D_1\right)$ and a \emph{diffusion operator} $\left(D_2\right)$. The diffusion term accounts for the smearing of the PDF due to process noise. Once the state PDF is computed through (\ref{FokkerPlanck}), the output PDF can again be obtained from (\ref{OutputPDF}).

\subsection{Discrete-time models}
\subsubsection{Uncertainty propagation for deterministic maps}
Let $\widehat{\mathcal{X}} \times \widehat{\mathcal{P}} \subseteq \mathbb{R}^{\widehat{n}_{s} + \widehat{n}_{p}}$ be a compact set and let $\mathcal{B}\left(\widehat{\mathcal{X}} \times \widehat{\mathcal{P}}\right)$ be the Borel-$\sigma$ algebra defined on it. Consider the discrete-time nonlinear system with state dynamics given by the vector recurrence relation
\begin{eqnarray}
\widehat{\widetilde{x}}_{k+1} = \widehat{\mathcal{T}}\left(\widehat{\widetilde{x}}_{k}\right), \; \widehat{\mathcal{T}}: \widehat{\mathcal{X}} \times \widehat{\mathcal{P}} \mapsto \widehat{\mathcal{X}} \times \widehat{\mathcal{P}},
\label{DeterministicMap}
\end{eqnarray}
where $\widehat{\mathcal{T}}$ is a measurable nonsingular transformation and the time index $k$ takes values from the ordered index set of non-negative integers $\{0, 1, 2, \hdots\}$. Then the evolution of the joint PDF $\widehat{\xi}\left(\widehat{\widetilde{x}}_{k}\right)$ is dictated by the \emph{Perron-Frobenius operator} $\widehat{\mathscr{P}}$, given by
\begin{eqnarray}
\displaystyle\int_{B} \widehat{\mathscr{P}} \widehat{\xi}\left(\widehat{\widetilde{x}}_{k}\right)\:d\widehat{\widetilde{x}}_{k} = \displaystyle\int_{\widehat{\mathcal{T}}^{-1}\left(B\right)} \widehat{\xi}\left(\widehat{\widetilde{x}}_{k}\right) \: d\widehat{\widetilde{x}}_{k}
\label{PFoperator}
\end{eqnarray}
for $B\in\mathcal{B}$. Properties of Perron-Frobenius operator can be found in Chap. 3 of \cite{LasotaMackey1994}. Further, assuming the output dynamics as $\widehat{y}_{k} = \widehat{h}\left(\widehat{\widetilde{x}}_{k}\right)$, one can derive $\widehat{\eta}\left(\widehat{y}_{k}\right)$ from $\widehat{\xi}\left(\widehat{\widetilde{x}}_{k}\right)$ using the discrete analogue of (\ref{OutputPDF}).

\subsubsection{Uncertainty propagation for stochastic maps}
In this case, we consider the nonlinear state space representation given by the stochastic maps of general form
\begin{eqnarray}
\widehat{\widetilde{x}}_{k+1} = \widehat{\mathcal{T}}\left(\widehat{\widetilde{x}}_{k}, \zeta_{k}\right), \qquad \widehat{\widetilde{y}}_{k} = \widehat{h}\left(\widehat{\widetilde{x}}_{k}, \zeta_{k}\right),
\label{StochasticMap}
\end{eqnarray}
where $\zeta_{k} \in \mathbb{R}^{\omega}$ is the i.i.d. sample drawn from a known distribution for the noise (stochastic perturbations). Here, the dynamics $\widehat{\mathcal{T}}$ is not required to be a non-singular transformation (Chap. 10, \cite{LasotaMackey1994}). Since $\widehat{\mathcal{T}}$ defines a Markov Chain on $\widehat{\mathcal{X}} \times \widehat{\mathcal{P}}$, it can be shown that \cite{LasotaMackey1994,MeynTweedie2009}  evolution of the joint PDFs follow
\begin{eqnarray}
\widehat{\xi}_{k+1} := \widehat{\xi}_{\widehat{\widetilde{x}}_{k+1}}\left(\widehat{\widetilde{x}}\right) = \displaystyle\int_{\widehat{\mathcal{X}} \times \widehat{\mathcal{P}}} \mathcal{K}_{\widehat{\mathcal{T}}}\left(\widehat{\widetilde{x}}|z\right) \xi_{\widehat{\widetilde{x}}_{k}}\left(z\right) dz, \nonumber \\
\widehat{\eta}_{k} := \widehat{\eta}_{\widehat{y}_{k}}\left(\widehat{y}\right) = \displaystyle\int_{\widehat{\mathcal{X}} \times \widehat{\mathcal{P}}} \mathcal{K}_{\widehat{h}}\left(\widehat{y}|z\right) \widehat{\xi}_{\widehat{\widetilde{x}}_{k}}\left(z\right) dz,
\label{StocMapPDFupdate}
\end{eqnarray}
where $\mathcal{K}_{\mathcal{T}}\left(\widehat{\widetilde{x}}|z\right)$ and $\mathcal{K}_{h}\left(\widehat{y}|z\right)$ are the \emph{stochastic kernels} for maps $\widehat{\mathcal{T}}$ and $\widehat{h}$ respectively. (\ref{StocMapPDFupdate}) can be seen as a special case of the Chapman-Kolmogorov equation \cite{Papoulis1984}.

\subsection{Computational aspects}
For deterministic flow, the Liouville PDE (\ref{Liouville}) can be solved in exact arithmetic \cite{HalderBhattacharya2011} via method-of-characteristics (MOC). Since the characteristic curves for (\ref{Liouville}) are the trajectories in the extended state space, $\widehat{\xi}\left(\widehat{\widetilde{x}},t\right)$ and hence $\widehat{\eta}\left(\widehat{y},t\right)$ can be computed directly along these characteristics. Unlike Monte-Carlo, this is an ``on-the-fly" computation and does not involve any approximation, and hence offers a superior performance \cite{HalderBhattacharya2011,HalderBhattacharya2010} than Monte-Carlo in high dimensions. For deterministic maps, cell-to-cell mapping \cite{Hsu1987} achieves a finite dimensional approximation of the Perron-Frobenius operator.

For stochastic flow, solving Fokker-Planck PDE (\ref{FokkerPlanck}) is numerically challenging \cite{Risken} but has seen some recent success \cite{Kumar2010} in moderate (4 to 5) dimensions. For high dimensional stochastic flows, an extension of the MOC approach has been proposed \cite{Dutta2011}. For stochastic maps, discretizations for stochastic kernels (11) and (12), can be done through cell-to-cell mapping \cite{Hsu1987} resulting a random transition probability matrix \cite{EdelmanSutton2007}.

%%%%%%%%%%%%%%%%%%%%%%%%%%%%%%%%%%%%%%%%%%%%%%%%%%%%%%%%%%%%%%%%%%%%%%%%%%%%%%%%%%%%%%%%%%%%%%%%%%%%%%%%%%%%%%%%%%%

\section{Distributional Comparison}
Once the observed and model-predicted output PDFs $\eta\left(y,t\right)$ and $\widehat{\eta}\left(\widehat{y},t\right)$, are obtained, we need a metric to compare the \emph{shapes} of these two PDFs at times $\{t_{j}\}_{j=1}^{\tau}$, when the measurement PDF $\eta\left(y,t_{j}\right)$ is available. We argue that the suitable metric for this purpose is \emph{Wasserstein distance}.

\subsection{Choice of metric}
Distances on the space of probability distributions \cite{GibbsSu2002}, can be broadly categorized into two classes, viz. Csis\'{z}ar's $\phi$-divergence \cite{ref5Perum} and integral probability metrics \cite{ref7Perum}. The first includes well-known distances like Kullback-Leibler (KL) divergence, Hellinger distance, $\chi^{2}$ divergence etc. while the latter includes Wasserstein distance, Dudley metric, maximum mean discrepancy. Total variation distance belongs to both of these classes.

The choice of a suitable metric depends on application. Following the intuitions of Section 2.1, we list the \emph{axiomatic requirements}, that a model validation metric must satisfy:

\begin{enumerate}
 \item [\textbf{R.1}] The notion of ``distance" must measure the \emph{shape difference} between two instantaneous output PDFs. This is because a good model must emulate similar concentration of trajectories as observed in the measurement space, i.e. the respective joint PDFs $\eta\left(y,t\right)$ and $\widehat{\eta}\left(\widehat{y},t\right)$, over the time-varying output supports, must match at times whenever measurements are available. In particular, the distance must be function of \emph{shape difference but not of shape}, i.e. same amount of shape difference must return same magnitude of distance, irrespective of the individual shapes being compared.\\

 \item [\textbf{R.2}] For meaningful validation inference, the choice of distance must be a metric.\\

 \item [\textbf{R.3}] For a given model-data pair, the supports of $\eta\left(y,t\right)$ and $\widehat{\eta}\left(\widehat{y},t\right)$ may not match at $t=t_{j}$, $j=1,\hdots,\tau$. The distance must be
     well defined and computable under such circumstances.\\

  \item [\textbf{R.4}] The computation of the distance need not require $\eta\left(y,t\right)$ and $\widehat{\eta}\left(\widehat{y},t\right)$ to be represented by the same number of samples. For the purpose of model validation, this offers practical advantages since experimental data are often expensive to gather. However, model based simulation can harness the computational resources and hence, simulation sample size is often larger than that of experimental data.\\

  \item [\textbf{R.5}] The distance must be asymptotically consistent with respect to finite sample representations of the PDFs under comparison. Namely, in the infinite sample limit, the empirical estimate of the distance must converge to the actual instantaneous value of the distance. For practical computation, this rate-of-convergence is required to be fast with respect to the number of samples.
 \end{enumerate}
Next, we introduce the Wasserstein distance on the manifold of PDFs, which will be shown to fulfil the axiomatic requirements listed above.

\begin{definition}
(\textbf{Wasserstein distance}) Let the $\ell_{p}$ norm between two random output vectors $y\in\mathcal{Y}\subseteq \mathbb{R}^{n_{o}}$, and $\widehat{y}\in\widehat{\mathcal{Y}}\subseteq \mathbb{R}^{n_{o}}$, be denoted as $\parallel y - \widehat{y}\parallel_{p}$. Then, the Wasserstein distance of order $q$, between two PDFs $\eta\left(y\right)$ and $\widehat{\eta}\left(\widehat{y}\right)$, is defined as
\begin{eqnarray}
_{p}W_{q}\left(\eta,\widehat{\eta}\right) := \left[\underset{\rho \in \mathcal{M}_{2}\left(\eta,\widehat{\eta}\right)} {\text{\emph{inf}}} \displaystyle\int_{\mathcal{Y}\times\widehat{\mathcal{Y}}} \|y-\widehat{y}\|_{p}^{q} \:\: \rho\left(y,\widehat{y}\right) \: dy d\widehat{y}\right]^{\frac{1}{q}} \;\,
\label{Wasserstein}
\end{eqnarray}
where $\mathcal{M}_{2}\left(\eta,\widehat{\eta}\right)$ is the set of all joint PDFs supported on $\mathcal{Y}\times\widehat{\mathcal{Y}}$, having finite second moments, with first marginal as $\eta$ and second marginal as $\widehat{\eta}$.
\label{WassDefn}
\end{definition}
\begin{remark} (\textbf{Generalizations})
In general, the sets $\mathcal{Y}$ and $\widehat{\mathcal{Y}}$ can be subsets of any complete, separable metric (Polish) space, equipped with a $p$\textsuperscript{th} order distance metric. Further, (\ref{Wasserstein}) does not require the distributions under comparison to be absolutely continuous. It remains well defined between output measures $\mu$ and $\widehat{\mu}$, even when the corresponding PDFs $\eta$ and $\widehat{\eta}$ don't exist.
\end{remark}
\begin{remark}(\textbf{Choice of $p=q=2$})
We take Euclidean metric ($p=2$) as the inter-sample distance between random vectors $y$ and $\widehat{y}$. Further, we set $q=2$ since it guarantees uniqueness \cite{Villani2003} in (\ref{Wasserstein}), and has the interpretation of minimum effort needed to morph a density shape to other. Also, Jordan, Kinderlehrer and Otto \cite{Jordan1998} have rigorously demonstrated that uncertainty propagation in a dynamical system can be seen as a gradient flux of free energy with respect to the Wasserstein distance of order $q=2$.
\end{remark}

The interpretation of $_{2}W_{2}$ as mass preserving optimal transport between two given shapes, makes it a strong candidate for model validation purpose. Further, it is known \cite{Rachev1991} that on the set $\mathcal{M}_{2}$, $_{2}W_{2}$ defines a metric. Thus, Wasserstein distance meets \textbf{R.1} and \textbf{R.2}. Also, \textbf{R.3} and \textbf{R.4} are satisfied since Definition \ref{WassDefn} does not require the supports or cardinality of the sample representations of the PDFs to be the same. This will be illustrated further in Section 4.2, when we describe the computation of $_{2}W_{2}$ between two scattered point clouds with probability weights. For \textbf{R.5}, convergence of sample Wasserstein estimate to its true deterministic value, will be discussed in Section 4.3.1 (Theorem \ref{WassEstimateUpperBound}).

\subsubsection{Limitations of pointwise distances}

Commonly used information-theoretic distances like Kullback-Leibler divergence $D_{KL}\left(\eta\parallel\widehat{\eta}\right) \triangleq \mathbb{E}[\log(\eta/\widehat{\eta})]$, its symmetrized version $D_{KL}^{\text{symm}} \triangleq D_{KL}\left(\eta\parallel\widehat{\eta}\right) + D_{KL}\left(\widehat{\eta}\parallel\eta\right)$, are not metrics. On the other hand, Hellinger distance $H\left(\eta,\widehat{\eta}\right) \triangleq \frac{1}{\sqrt{2}} \parallel \sqrt{\eta} - \sqrt{\widehat{\eta}}\parallel_{L_{2}\left(\mathbb{R}^{n_{o}}\right)}$, and the square-root of Jensen-Shannon divergence $JSD\left(\eta,\widehat{\eta}\right) \triangleq \frac{1}{2}\left[D_{KL}\left(\eta\parallel\frac{1}{2}\left(\eta+\widehat{\eta}\right)\right) + D_{KL}\left(\widehat{\eta}\parallel\frac{1}{2}\left(\eta+\right.\right.\right.$ $\left.\left.\left.\widehat{\eta}\right)\right)\right]$ are metrics. However, being pointwise definitions, all of them fail to satisfy \textbf{R.3} and \textbf{R.4}, resulting computational difficulties for model validation. As for \textbf{R.5}, $D_{KL}\left(\eta\parallel\widehat{\eta}\right)$ is known to be asymptotically consistent, but the rate-of-convergence can be arbitrarily slow \cite{WangKulkarniVerdu2005,NguyenWainwrightJordan2010}. Besides these computational problems, we emphasize here that the information theoretic distances may not discriminate shapes in a geometric sense, as desired in \textbf{R.1}. We provide two counterexamples below to illustrate this point. The first counterexample highlights that two PDFs with same randomness need not have similar shapes. The second counterexample demonstrates that $D_{KL}$ may depend on the shapes under comparison.

\begin{figure*}[!htb]
\vspace*{-0.1in}
\minipage{0.32\textwidth}
\includegraphics[width=\linewidth]{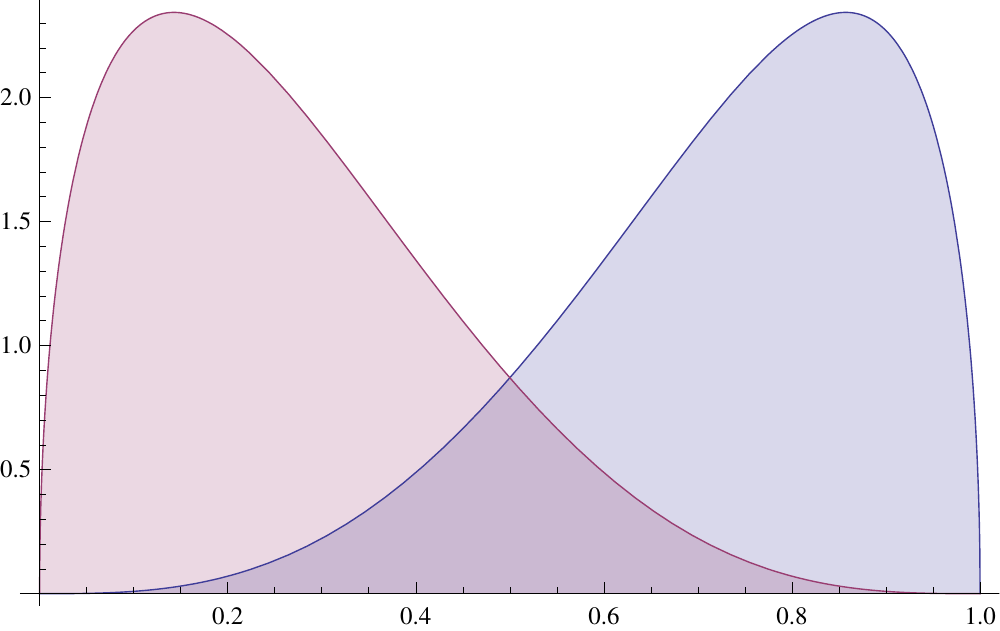}
\caption{\small{The two beta densities $\eta_{b}\left(x;\alpha,\beta\right)$ (left-skewed) and $\eta_{b}\left(x;\beta,\alpha\right)$ (right-skewed) with $\alpha = 4$, $\beta=\frac{3}{2}$, have same entropy/randomness, but have different shapes.}}
\label{BetaSymmeric}
\endminipage\hfill
\minipage{0.32\textwidth}
\includegraphics[width=\linewidth]{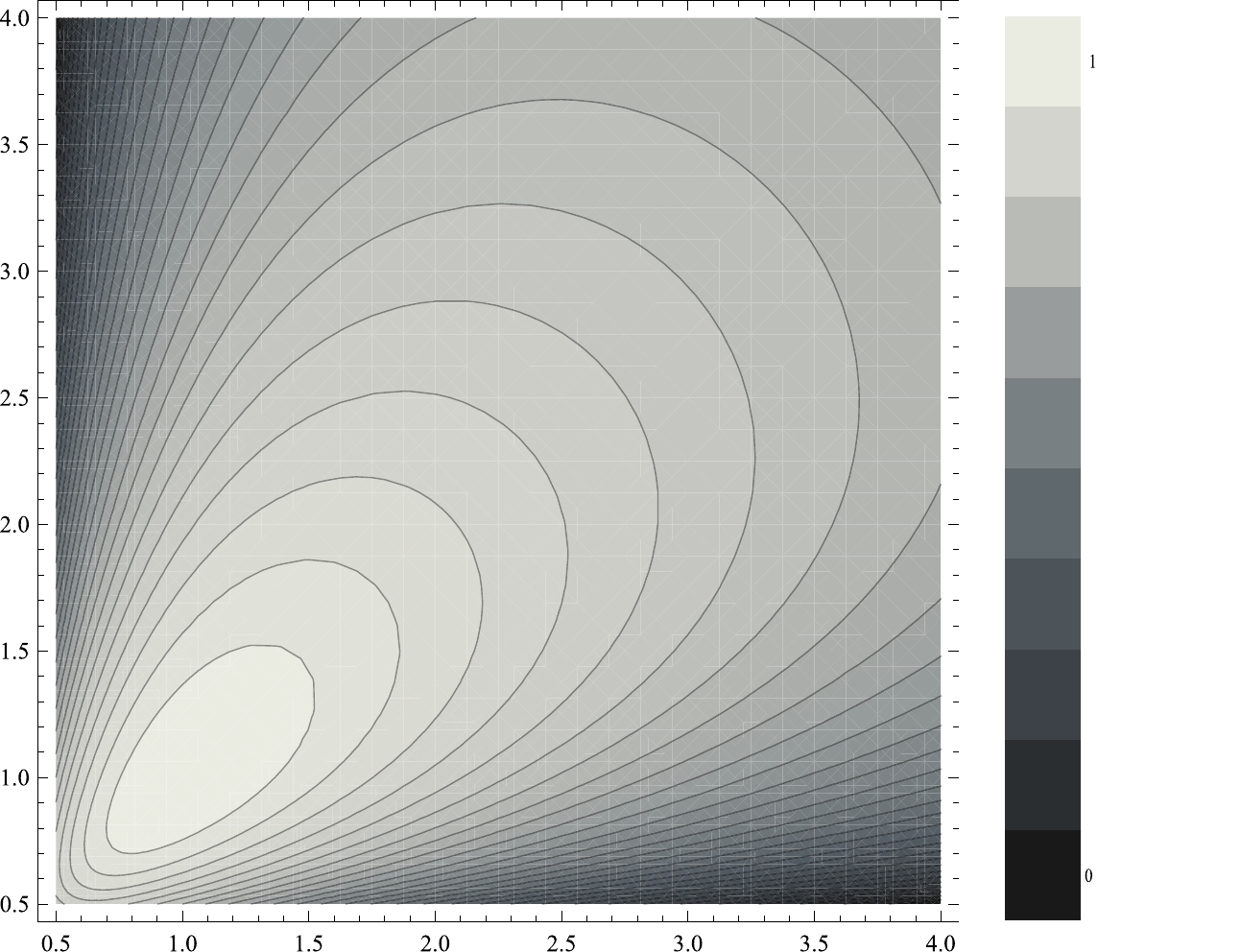}
\caption{\small{Isentropic contours of beta family in $\left(\alpha,\beta\right)$ space. The symmetry of the contours about $\alpha=\beta$ line implies $H_{b}(\alpha,\beta) = H_{b}(\beta,\alpha)$. This plot also shows that uniform distribution $\left(\alpha=\beta=1\right)$ is of maximum entropy.}}
\label{IsoEntropyBetaFamily}
\endminipage\hfill
\minipage{0.32\textwidth}
\vspace*{-0.65in}
\includegraphics[width=\linewidth]{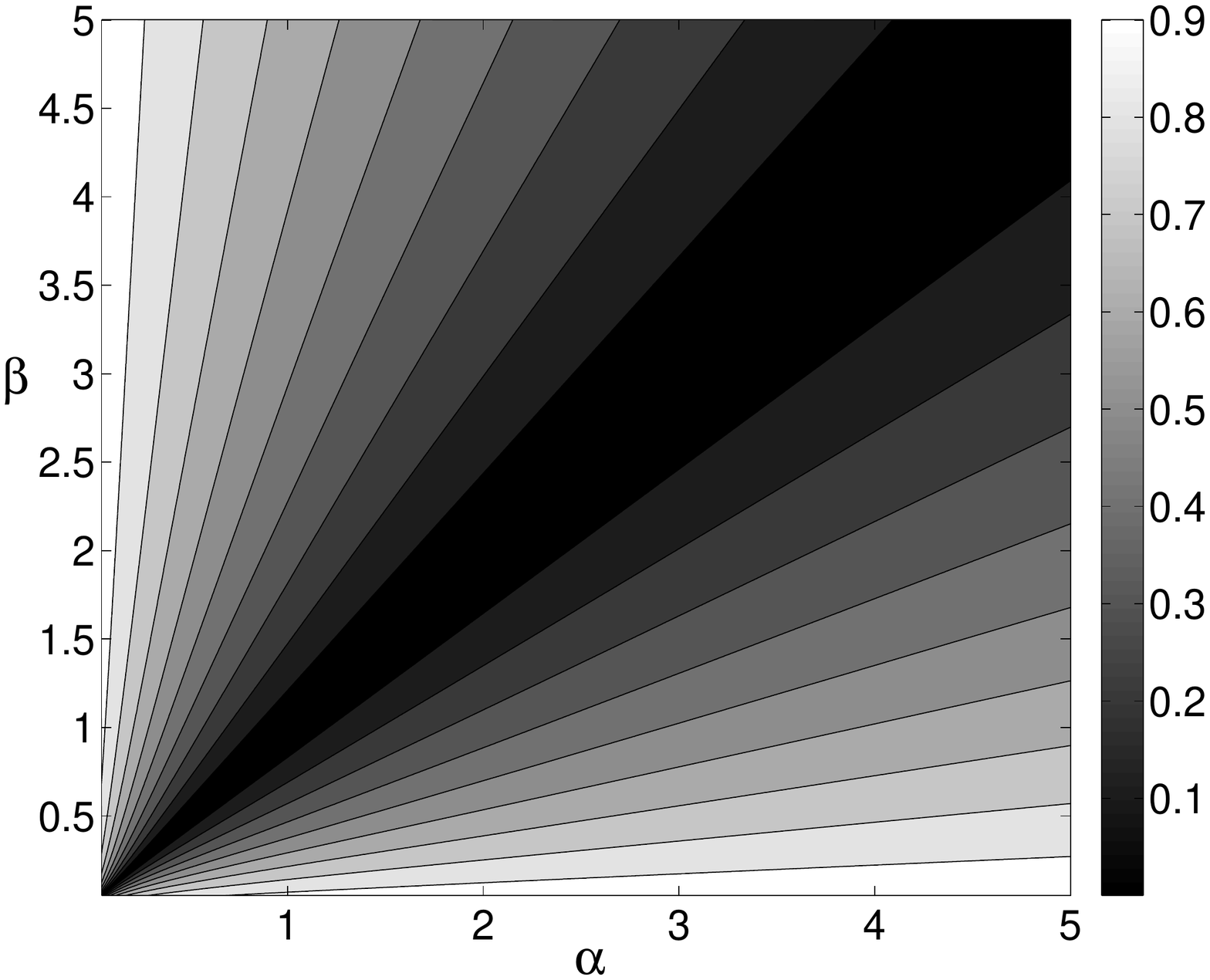}
\vspace*{-0.75in}
\caption{\small{Iso-Wasserstein contours of $_{2}W_{2}\left(\eta_{b}\left(x;\alpha,\beta\right), \eta_{b}\left(x;\beta,\alpha\right)\right)$ in $\left(\alpha,\beta\right)$ space. Since $_{2}W_{2}$ is a metric, it has symmetry about $\alpha = \beta$ line, and vanishes only along this line. The computation of $_{2}W_{2}$ is detailed in Appendix A.}}
\label{SomethingElse}
\endminipage
\end{figure*}

\begin{counterexample}(\textbf{Randomness $\neq$ shape})
Consider the two parametric family of beta densities $\eta_{b}\left(x;\alpha,\beta\right) \triangleq \frac{x^{\alpha-1} \left(1-x\right)^{\beta-1}}{B\left(\alpha,\beta\right)}$, $\alpha, \beta > 0$, $x \in \left[0, 1\right]$, where $B\left(\alpha,\beta\right) \triangleq \int_{0}^{1} t^{\alpha-1} \left(1-t\right)^{\beta-1}\: dt = \frac{\Gamma\left(\alpha\right)\Gamma\left(\beta\right)}{\Gamma\left(\alpha + \beta\right)}$, is the complete beta function, and $\Gamma\left(z\right)$ denotes the gamma function. The differential entropy for beta family can be computed as \cite{LazoRathie1978}
\begin{eqnarray}
&&H_{b}\left(\alpha,\beta\right) = - \displaystyle\int_{0}^{1} \eta_{b}\left(x;\alpha,\beta\right) \log \eta_{b}\left(x;\alpha,\beta\right) \: dx \nonumber\\
&=& \log B\left(\alpha, \beta\right) - \left(\alpha - 1\right) \left(\Psi\left(\alpha\right) - \Psi\left(\alpha+\beta\right)\right) \nonumber\\
&& - \left(\beta - 1\right) \left(\Psi\left(\beta\right) - \Psi\left(\alpha+\beta\right)\right),
\label{BetaEntropy}
\end{eqnarray}
where $\Psi\left(z\right) \triangleq \frac{d}{dz} \log \Gamma\left(z\right)$, is the digamma function. Since (\ref{BetaEntropy}) remains invariant under $\left(\alpha,\beta\right) \mapsto \left(\beta,\alpha\right)$, $\alpha \neq \beta$, $\eta_{b}\left(x;\alpha,\beta\right)$ and $\eta_{b}\left(x;\beta,\alpha\right)$ have same entropy, but one is skewed to right and the other to left, as shown in Fig. \ref{BetaSymmeric}. Fig. \ref{IsoEntropyBetaFamily} shows the isentropic contours of beta PDFs in $\left(\alpha,\beta\right)$ space. Any pair of \textbf{distinct} points chosen on these contours, results two beta PDFs with non-identical shapes, as revealed by Fig. \ref{SomethingElse} and Appendix A.
\end{counterexample}

\begin{counterexample}(\textbf{$D_{KL} \neq$ shape difference})
Consider two $\nu$-dimensional homoscedastic Gaussian PDFs $\mathcal{N}\left(m_{1},\Sigma_{1}\right)$ and $\mathcal{N}\left(m_{2},\Sigma_{2}\right)$, such that $\Sigma_{1} = \Sigma_{2}$. Since the only \textbf{difference} between the two PDFs is the location of their means, a shape-discriminating distance is expected to be a function of $\parallel m_{1} - m_{2} \parallel_{2}$, and should not depend on the covariance matrix i.e. \textbf{shapes of the individual PDFs}.

In this situation, $_{2}W_{2} = \Vert m_{1} - m_{2} \Vert_{2}$ \cite{GivensShortt1984} and $D_{KL} = \displaystyle\frac{1}{2}\left(m_{2} - m_{1}\right)^{\top} \Sigma_{2}^{-1} \left(m_{2} - m_{1}\right)$ \cite{Kullhavy1996}. If we introduce $m := m_{2} - m_{1}$, then $\displaystyle\frac{D_{KL}}{_{2}W_{2}} = \displaystyle\frac{\Vert m \Vert_{2}}{2} \: r$, where $r := \displaystyle\frac{m^{\top} \Sigma_{2}^{-1} m}{m^{\top} m}$ is the Rayleigh quotient corresponding to the positive semi-definite precision matrix $\Sigma_{2}^{-1}$. It's known (Chap. 7, \cite{Poznyak2008}) that if we denote
$\mathcal{K} := \{\lambda : \lambda = \displaystyle\sum_{i=1}^{\nu} \alpha_{i}\lambda_{i}, \: \displaystyle\sum_{i=1}^{\nu} \alpha_{i} = 1, \: \alpha_{i} \geqslant 0, \: \forall i =1, 2, \hdots, \nu \}$ as the convex hull of the eigenvalues of the precision matrix $\Sigma_{2}^{-1}$, then $r\left(m\right) \in \mathcal{K}$. In particular,
\begin{eqnarray*}
r_{\text{min}} = \lambda_{\text{min}}\left(\Sigma_{2}^{-1}\right) = \displaystyle\frac{1}{\lambda_{\text{min}}\left(\Sigma_{2}\right)} > 0, \nonumber\\
r_{\text{max}} = \lambda_{\text{max}}\left(\Sigma_{2}^{-1}\right) = \displaystyle\frac{1}{\lambda_{\text{max}}\left(\Sigma_{2}\right)} > 0,
\end{eqnarray*}
and these extrema are attained when $m:= m_{2} - m_{1}$ respectively coincides with the minimum and maximum eigenvector of $\Sigma_{2}^{-1}$. Thus the spectrum of $\Sigma_{2}^{-1}$ governs the magnitude of the ratio $\displaystyle\frac{D_{KL}}{_{2}W_{2}}$, even when $\Vert m \Vert_{2}$ is kept fixed. In particular, the ratio assumes unity iff $r = \displaystyle\frac{2}{\Vert m \Vert_{2}} \Rightarrow \Sigma_{2}^{-1} = \displaystyle\frac{2}{\Vert m \Vert_{2}} \: I_{\nu} \Rightarrow \Sigma_{1} = \Sigma_{2} = \displaystyle\frac{\Vert m \Vert_{2}}{2} \: I_{\nu}$.

Further discussions on the inadequacy of $D_{KL}$ for capturing shape characteristics and the utility of Wasserstein distance for the same, can be found in \cite{Hong2008,Ni2009}.
\end{counterexample}

\subsubsection{Wasserstein gap between dynamical systems}

\begin{proposition}(\textbf{Single output systems})\cite{Vallander1973}
 At time $t>0$, let $F\left(y,t\right)$ and $\widehat{F}\left(\widehat{y},t\right)$ be the cumulative distribution functions (CDFs) corresponding to the univariate PDFs $\eta\left(y,t\right)$ and $\widehat{\eta}\left(\widehat{y},t\right)$, respectively. Then
 \begin{eqnarray}
 _{2}W_{2}\left(t\right) &=& \sqrt{\displaystyle\int_{0}^{1} \left(F^{-1}\left(\varsigma,t\right) - \widehat{F}^{-1}\left(\varsigma,t\right)\right)^{2} \: d\varsigma}, \\
 \rho^{\star}\left(y,\widehat{y},t\right) &=& \min\left(F\left(y,t\right), \widehat{F}\left(\widehat{y},t\right)\right),
 \label{SingleOutputW}
 \end{eqnarray}
where $\rho^{\star}$ is the optimizer in (\ref{Wasserstein}).
\end{proposition}
\begin{proposition}(\textbf{Linear Gaussian systems})
Consider stable, observable LTI system pairs in continuous and discrete time:
\begin{eqnarray}
dx_{i}(t) = A_{i} x_{i}(t) dt + B_{i} d\beta_{i}(t), \quad y_{i}(t) = C_{i} x_{i}(t), \\
x_{i}(k+1) = A_{i} x_{i}(k) + B_{i} \vartheta_{i}(k), \quad y_{i}(k) = C_{i} x_{i}(k),
\end{eqnarray}
where $i=1,2$. $\beta_{i}(t)$ are Wiener processes with auto-covariances $Q_{i}\left(t_{1} \wedge t_{2}\right)$, $t_{1}, t_{2} > 0$, and $\vartheta_{i}\left(k\right)$ are Gaussian white noises with covariances $Q_{i}\left(k\right)$. If the initial PDF $\xi_{0} = \mathcal{N}\left(\mu_{0},\Sigma_{0}\right)$, then the Wasserstein distance between output PDFs $\eta_{i} = \mathcal{N}\left(\mu_{y_{i}},\Sigma_{y_{i}}\right)$, is given by \cite{GivensShortt1984}
\vspace*{-0.2in}
{\small{\begin{eqnarray}
_{2}W_{2} = \sqrt{\parallel\mu_{y_{1}} - \mu_{y_{2}}\parallel_{2}^{2} + \text{\emph{tr}}\left(\Sigma_{y_{1}} + \Sigma_{y_{2}} - 2 \left[\sqrt{\Sigma_{y_{1}}} \Sigma_{y_{2}} \sqrt{\Sigma_{y_{1}}}\right]^{\frac{1}{2}}\right)},
\label{GaussianWass}
\end{eqnarray}}}
\noindent
where $\mu_{y_{i}} = C_{i} \mu_{x_{i}}$, $\Sigma_{y_{i}} = C_{i} \Sigma_{x_{i}} C_{i}^{\top}$. For the continuous-time case,
\begin{eqnarray}
\dot{\mu}_{x_{i}}(t) &=& A_{i} \mu_{x_{i}}(t), \\
\dot{\Sigma}_{x_{i}}(t) &=& A_{i}\Sigma_{x_{i}}(t) + \Sigma_{x_{i}}(t) A_{i}^{\top} + B_{i} Q_{i} B_{i}^{\top},
\label{ContTimeMuSigma}
\end{eqnarray}
and for the discrete-time case,
\begin{eqnarray}
\mu_{x_{i}}\left(k+1\right) &=& A_{i} \mu_{x_{i}}\left(k\right), \\
\Sigma_{x_{i}}\left(k+1\right) &=& A_{i} \Sigma_{x_{i}}\left(k\right) A_{i}^{\top} + B_{i} Q_{i} B_{i}^{\top},
\label{DiscreteTimeMuSigma}
\end{eqnarray}
to be solved with $\mu_{x_{i}}\left(0\right) = \mu_{0}$, and $\Sigma_{x_{i}}\left(0\right) = \Sigma_{0}$. Deterministic results are recovered from above by setting the diffusion matrix $B_{i} = 0$.
\end{proposition}

\begin{remark}
(\textbf{Asymptotic Wasserstein distance})
In Table \ref{WassTable}, we have listed asymptotic Wasserstein distances between different pairs of stable dynamical systems. The asymptotic $_{2}W_{2}$ between two deterministic linear systems (\textbf{first row}) is zero since the origin being unique equilibria for both systems, Dirac delta is the stationary density for both. For a pair of deterministic affine systems (\textbf{second row}), asymptotic $_{2}W_{2}$ is simply the $\ell_{2}$ norm between their respective fixed points. This holds true even for a pair of nonlinear systems, each having a \textbf{unique} globally asymptotically stable equilibrium. For the stochastic linear case (\textbf{third row}), $\Sigma_{y\infty} = C \Sigma_{x\infty} C^{\top}$, and $\widehat{\Sigma}_{\widehat{y}\infty} = \widehat{C} \widehat{\Sigma}_{\widehat{x}\infty} \widehat{C}^{\top}$; where $\Sigma_{x\infty}, \widehat{\Sigma}_{\widehat{x}\infty}$ respectively solve $A \Sigma_{x\infty} + \Sigma_{x\infty} A^{\top} + B Q B^{\top} = 0$, and $\widehat{A} \widehat{\Sigma}_{\widehat{x}\infty} + \widehat{\Sigma}_{\widehat{x}\infty} \widehat{A}^{\top} + \widehat{B} \widehat{Q} \widehat{B}^{\top} = 0$. $Q$ and $\widehat{Q}$ are process noise covariances associated with Wiener processes $\beta\left(t\right)$ and $\widehat{\beta}\left(t\right)$. For the \textbf{fourth} and \textbf{fifth row}, the set of stable equilibria for the true and model nonlinear system, are given by $\{y_{i}^{\star}\}_{i=1}^{n^{\star}}$ and $\{\widehat{y}_{i}^{\star}\}_{i=1}^{\widehat{n}^{\star}}$, respectively. Further, we assume that the nonlinear systems have no invariant sets other than these stable equilibria. In such cases, the stationary densities are convex sum of Dirac delta densities, located at these equilibria. The weights for this convex sum, denoted as $m_{i}^{\star}$ and $\widehat{m}_{i}^{\star}$, depend on the initial PDF $\xi_{0}$. In particular, if we denote $\mathcal{R}_{i}$ as the \textbf{region-of-attraction} of the $i$\textsuperscript{th} equilibrium, then (see Appendix B)
\begin{eqnarray}
m_{i}^{\star} = \displaystyle\int_{\text{\emph{supp}}\left(\xi_{0}\right) \cap \mathcal{R}_{i}} \xi_{0}\left(x_{0}\right) \: dx_{0} \; \in \: \left[0,1\right].
\label{MassFractionFormula}
\end{eqnarray}
 To further illustrate this idea, a numerical example corresponding to the \textbf{fourth row} in Table \ref{WassTable}, will be provided in Section 6.
\label{MassFractionRemark}
\end{remark}

\begin{table*}
\caption{For various stable dynamical system pairs, we list asymptotic Wasserstein distance, defined as $_{2}W_{2}\left(\eta_{\infty},\widehat{\eta}_{\infty}\right)$, where $\eta_{\infty}$ and $\widehat{\eta}_{\infty}$ are the stationary PDFs of the true and model dynamics, respectively.}
\begin{center}
\tabcolsep 1pt
\scriptsize
\begin{tabular}{|c|c|c|c|}\hline \hline
Systems & Dynamics & Stationary PDFs & Asymptotic $_{2}W_{2}$\\ \hline \hline
Deterministic linear pair & $\dot{x}(t) = A x(t), \: y(t) = Cx(t)$, & $\eta_{\infty} = \delta\left(y\right)$ & 0\\
 & $\dot{\widehat{x}}(t) = \widehat{A} \widehat{x}(t), \: \widehat{y}(t) = \widehat{C}\widehat{x}(t)$ & $\widehat{\eta}_{\infty} = \delta\left(\widehat{y}\right)$ & \\ \hline
Deterministic affine pair & $\dot{x}(t) = A x(t) + b, \: y(t) = Cx(t) + d$, & $\eta_{\infty} = \delta\left(y + CA^{-1}b - d\right)$ & $\bigg\lVert \left(d-\widehat{d}\right) - \left(CA^{-1}b - \widehat{C}\widehat{A}^{-1}\widehat{b}\right)\bigg\rVert_{2}$\\
 & $\dot{\widehat{x}}(t) = \widehat{A} \widehat{x}(t) + \widehat{b}, \: \widehat{y}(t) = \widehat{C}\widehat{x}(t) + \widehat{d}$ & $\widehat{\eta}_{\infty} = \delta\left(\widehat{y} + \widehat{C}\widehat{A}^{-1}\widehat{b} - \widehat{d}\right)$ & \\ \hline
Stochastic linear pair & $dx(t) = A x(t) dt + B d\beta(t), \: y(t) = Cx(t)$, & $\eta_{\infty} = \mathcal{N}\left(0,\Sigma_{y\infty}\right)$ & $\left(\text{tr}\left(\Sigma_{y\infty} + \widehat{\Sigma}_{\widehat{y}\infty} - 2 \left[\Sigma_{y\infty}^{\frac{1}{2}}\widehat{\Sigma}_{\widehat{y}\infty}\Sigma_{y\infty}^{\frac{1}{2}}\right]^{\frac{1}{2}}\right)\right)^{\frac{1}{2}}$\\
  & $d\widehat{x}(t) = \widehat{A} \widehat{x}(t) dt + \widehat{B} d\widehat{\beta}(t), \: \widehat{y}(t) = \widehat{C}\widehat{x}(t)$ & $\widehat{\eta}_{\infty} = \mathcal{N}\left(0,\widehat{\Sigma}_{\widehat{y}\infty}\right)$ & \\ \hline
Deterministic nonlinear & $\dot{x}(t) = f\left(x(t)\right), \: y(t) = h\left(x(t)\right)$, & $\eta_{\infty} = \displaystyle\sum_{i=1}^{n^{\star}} m_{i}^{\star} \delta\left(y - y_{i}^{\star}\right)$ & $\left(\displaystyle\sum_{i=1}^{n^{\star}} \big\lVert y_{i}^{\star} \big\rVert_{2}^{2} \: m_{i}^{\star}\delta\left(y - y_{i}^{\star}\right)\right)^{\frac{1}{2}}$\\
 and deterministic linear & $\dot{\widehat{x}}(t) = \widehat{A} \widehat{x}(t), \: \widehat{y}(t) = \widehat{C}\widehat{x}(t)$ & $\widehat{\eta}_{\infty} = \delta\left(\widehat{y}\right)$ & \\ \hline
 Deterministic nonlinear pair & $\dot{x}(t) = f\left(x(t)\right), \: y(t) = h\left(x(t)\right)$, & $\eta_{\infty} = \displaystyle\sum_{i=1}^{n^{\star}} m_{i}^{\star} \delta\left(y - y_{i}^{\star}\right)$ & Monge-Kantorovich optimal\\
 & $\dot{\widehat{x}}(t) = \widehat{f}\left(\widehat{x}(t)\right), \: \widehat{y}(t) = \widehat{h}\left(\widehat{x}(t)\right)$ & $\widehat{\eta}_{\infty} = \displaystyle\sum_{i=1}^{\widehat{n}^{\star}} \widehat{m}_{i}^{\star} \delta\left(\widehat{y} - \widehat{y}_{i}^{\star}\right)$ & transport LP (\ref{HitchcockKoopmansLP}), (C1)--(C3)\\ \hline \hline
\end{tabular}
\end{center}
\label{WassTable}
\end{table*}

\subsection{Computing multivariate $_{2}W_{2}$}
Computing Wasserstein distance from (\ref{Wasserstein}) calls for solving \emph{Monge-Kantorovich optimal transportation plan} \cite{Rachev1985}. In this formulation, the difference in shape between two statistical distributions is quantified by the minimum amount of work required to convert a shape to the other. The ensuing optimization, often known as \emph{Hitchcock-Koopmans problem} \cite{Hitchcock1941,Koopmans1949,Koopmans1951}, can be cast as a linear program (LP), as described next.

Consider a complete, weighted, directed bipartite graph $K_{m,n}\left(U \cup V, E\right)$ with $\#\left(U\right) = m$ and $\#\left(V\right) = n$. If $u_{i} \in U, i=1,\hdots,m$, and $v_{j}\in V, j=1,\hdots,n$, then the edge weight $c_{ij}:=\parallel u_{i} - v_{j} \parallel_{\ell_{2}}^{2}$ denotes the cost of transporting unit mass from vertex $u_{i}$ to $v_{j}$. Then, according to (\ref{Wasserstein}), computing $_{2}W_{2}^{2}$ translates to
\begin{eqnarray}
\text{minimize} \; \displaystyle\sum_{i=1}^{m}\displaystyle\sum_{j=1}^{n} \: c_{ij} \: \varphi_{ij}
\label{HitchcockKoopmansLP}
\end{eqnarray}
subject to the constraints
\begin{equation*}
\displaystyle\sum_{j=1}^{n} \varphi_{ij} = \alpha_{i}, \qquad \forall \; u_{i} \in U,
\tag{C1}
\end{equation*}
\begin{equation*}
\displaystyle\sum_{i=1}^{m} \varphi_{ij} = \beta_{j}, \qquad \forall \; v_{j} \in V,
\tag{C2}
\end{equation*}
\begin{equation*}
\qquad\qquad\qquad\;\varphi_{ij} \geqslant 0, \qquad\quad \forall \; \left(u_{i},v_{j}\right) \in U \times V.
\tag{C3}
\end{equation*}
The objective of (\ref{HitchcockKoopmansLP}) is to come up with an optimal mass transportation policy $\varphi_{ij} := \varphi\left(u_{i} \rightarrow v_{j}\right)$ associated with cost $c_{ij}$. Clearly, in addition to constraints (C1)--(C3), (\ref{HitchcockKoopmansLP}) must respect the necessary feasibility condition
\begin{equation*}
\displaystyle\sum_{i=1}^{m} \alpha_{i} = \displaystyle\sum_{j=1}^{n} \beta_{j}
\tag{C0}
\end{equation*}
denoting the conservation of mass. In our context of measuring the shape difference between two PDFs, we treat the joint probability mass function (PMF) vectors $\alpha_{i}$ and $\beta_{j}$ to be the marginals of some unknown joint PMF $\varphi_{ij}$ supported over the product space $U \times V$. Since determining joint PMF with given marginals is not unique, (\ref{HitchcockKoopmansLP}) strives to find that particular joint PMF which minimizes the total cost for transporting the probability mass while respecting the normality condition. Notice that the finite-dimensional LP (\ref{HitchcockKoopmansLP}) is a direct discretization of the Wasserstein definition (\ref{Wasserstein}), and it is known \cite{sriperumbudur2012} that the solution of (\ref{HitchcockKoopmansLP}) is asymptotically consistent with that of the infinite dimensional LP (\ref{Wasserstein}).

\subsection{Computational complexity for $_{2}W_{2}$}

\subsubsection{Sample complexity}
For a desired accuracy of Wasserstein distance computation, we want to specify the bounds for number of samples $m=n$, for a given initial PDF. Since the finite sample estimate of Wasserstein distance is a random variable, we need to answer how large should $n$ be, in order to guarantee that the empirical estimate of Wasserstein distance obtained by solving the LP (\ref{HitchcockKoopmansLP}), (C1)--(C3) with $m=n$, is close to the true deterministic value of (\ref{Wasserstein}) in probability. In other words, given $\epsilon, \delta \in \left(0,1\right)$, we want to estimate a lower bound of $m=n$ as a function of $\epsilon$ and $\delta$, such that
\begin{eqnarray*}
\mathbb{P}\left(\left\lvert _{2}W_{2}\left(\eta_{m}^{j}\left(y\right), \widehat{\eta}_{n}^{j}\left(\widehat{y}\right)\right) \: - \: _{2}W_{2}\left(\eta^{j}\left(y\right), \widehat{\eta}^{j}\left(\widehat{y}\right)\right)\right\lvert < \epsilon\right) \\
> 1 - \delta, \qquad \forall j = 1, 2, \hdots, \tau.
\end{eqnarray*}
Similar consistency and sample complexity results are available in the literature (see Corollary 9(i) and Corollary 12(i) in \cite{Sriperumbudur2009}) for Wasserstein distance of order $q=1$. From H\"{o}lder's inequality, $W_{q_{2}} > W_{q_{1}}$ for $q_{2} > q_{1}$, and hence that sample complexity bound, in general, does not hold for $q=2$.To proceed, we need the following results.
 \begin{lemma}
(Appendix C) Given random variables $X$, $Y$, $Z$, such that $X \leqslant Y + Z$, then for $\epsilon > 0$, we have
\vspace*{-0.2in}
{\small{\begin{eqnarray*}
\mathbb{P}\left(X > \epsilon\right) \: \leqslant \: \mathbb{P}\left(Y + Z > \epsilon\right) \: \leqslant \: \mathbb{P}\left(Y > \frac{\epsilon}{2}\right) +\mathbb{P}\left(Z > \frac{\epsilon}{2}\right).
\end{eqnarray*}}}
\label{RandomVariableInequality}
\end{lemma}
\vspace*{-0.2in}
\begin{definition} (\textbf{Transportation cost inequality})\cite{Talagrand1996}
A probability measure $\mu$ is said to satisfy the $L_{p}$-\emph{transportation cost inequality} (TCI) of order $q$, if there exists some constant $C > 0$ such that for any probability measure $\nu$, $_{p}W_{q}\left(\mu, \nu\right) \leqslant \sqrt{2 C D_{KL}\left(\nu \parallel \mu\right)}$. In short, we write $\mu \in T_{q}\left(C\right)$. In particular, for $\mu \sim \mathcal{N}\left(m_{\kappa \times 1}, \Sigma_{\kappa \times \kappa}\right)$, we have  \cite{Djellout2004} $\mu \in T_{2}\left(\lambda_{\text{max}}\left(\Sigma\right)\right)$.
\end{definition}
\begin{theorem}
(\textbf{Rate-of-convergence of empirical measure in Wasserstein metric})(Thm. 5.3, \cite{BoissardleGouic2011})
For a probability measure $\rho \in T_{q}\left(\mathscr{C}\right)$, $1 \leqslant q \leqslant 2$, and its $n$-sample estimate $\rho_{n}$, we have
\begin{eqnarray}
\mathbb{P}\left( _{p}W_{q}\left(\rho, \rho_{n}\right) > \theta \right) \leqslant K_{\theta} \: \exp\left(-\displaystyle\frac{n\theta^{2}}{8\mathscr{C}}\right), \quad \theta > 0,
\label{ROCempiricalMeasure}
\end{eqnarray}
\noindent
 and $ \log K_{\theta} := \displaystyle\frac{1}{\mathscr{C}}\: \underset{\mu}{\text{\emph{inf}}}\, \#\left(\text{\emph{supp}} \: \mu\right) \left(\text{\emph{diam}}\left(\text{\emph{supp}} \: \mu\right)\right)^{2}$. The optimization takes place over all probability measures $\mu$ of finite support, such that $_{p}W_{q}\left(\rho,\mu\right) \leqslant \theta/4$.
\label{ROCempiricalMeasure}
\end{theorem}
 We now make few notational simplifications. In this subsection, we denote $\eta^{j}\left(y\right)$ and $\widehat{\eta}^{j}\left(y\right)$ by $\eta$ and $\widehat{\eta}$, and their finite sample representations by $\eta_{m}$ and $\widehat{\eta}_{n}$, respectively. Then we have the following result.
 \begin{theorem}
(\textbf{Rate-of-convergence of empirical Wasserstein estimate})
(Appendix D) For true densities $\eta$ and $\widehat{\eta}$, let corresponding empirical densities be $\eta_{m}$ and $\widehat{\eta}_{n}$, evaluated at respective uniform sampling of cardinality $m$ and $n$. Let $\mathscr{C}_{1}$, $\mathscr{C}_{2}$, be the TCI constants for $\eta$ and $\widehat{\eta}$, respectively and fix $\epsilon>0$. Then
\begin{eqnarray}
\mathbb{P}\left(\bigg\lvert \: _{2}W_{2}\left(\eta_{m}, \widehat{\eta}_{n}\right) \: - \: _{2}W_{2}\left(\eta, \widehat{\eta}\right) \bigg\rvert > \epsilon\right) \nonumber\\
\leqslant K_{1} \: \exp\left(-\displaystyle\frac{m \epsilon^{2}}{32 \mathscr{C}_{1}}\right)
+ K_{2} \: \exp\left(-\displaystyle\frac{n \epsilon^{2}}{32 \mathscr{C}_{2}}\right).
\label{OurWassEstimate}
\end{eqnarray}
\label{WassEstimateUpperBound}
\end{theorem}
\begin{remark}
At a fixed time, $K_{1}$, $K_{2}$, $\mathscr{C}_{1}$ and $\mathscr{C}_{2}$ are constants in a given model validation problem, i.e. for a given pair of experimental data and proposed model. However, values of these constants depend on true and model dynamics. In particular, the TCI constants $\mathscr{C}_{1}$ and $\mathscr{C}_{2}$ depend on the dynamics via respective PDF evolution operators. The constants $K_{1}$ and $K_{2}$ depend on $\eta$ and $\widehat{\eta}$, which in turn depend on the dynamics. For pedagogical purpose, we next illustrate the simplifying case $K_{1} = K_{2} = K$, $\mathscr{C}_{1} = \mathscr{C}_{2} = \mathscr{C}$.
\end{remark}
\begin{corollary}
(\textbf{Sample complexity for empirical Wasserstein estimate})
For desired accuracy $\epsilon \in \left(0, 1\right)$, and confidence $1 - \delta$, $\delta \in \left(0,1\right)$, the sample complexity $m = n = N_{\text{wass}}$, for finite sample Wasserstein computation is given by
\begin{eqnarray}
N_{\text{wass}} = \left(\displaystyle\frac{32 \mathscr{C}}{\epsilon^{2}}\right) \: \log\left(\displaystyle\frac{2 K}{\delta}\right).
\label{SampleComplexityWass}
\end{eqnarray}
\end{corollary}

\subsubsection{Runtime complexity}
The LP formulation (\ref{HitchcockKoopmansLP}), (C1)--(C3), requires solving for $mn$ unknowns subject to $\left(m + n + mn\right)$ constraints. For $m=n$, it can be shown that \cite{Burkard2009,Julien2010} the runtime complexity for solving the LP is $O\left(n_{o} \: n^{2.5}\log\nu\right)$. Notice that the output dimension $n_{o}$ enters only through the cost $c_{ij}$ in (\ref{HitchcockKoopmansLP}) and hence affects the computational time linearly.

In actual simulations, we found the runtime of the LP (\ref{HitchcockKoopmansLP}) to be sensitive on how the constraints were implemented. Suppose, we put (\ref{HitchcockKoopmansLP}) in standard form
\begin{eqnarray}
\text{minimize} \; \widetilde{c}^{\top} \widetilde{\varphi}, \qquad \text{subject to} \; A\widetilde{\varphi} = b, \quad \widetilde{\varphi} \geqslant 0,
\label{StandardLPform}
\end{eqnarray}
where $\widetilde{c}_{mn \times 1} := \text{vec}\left(c\right)$, $\widetilde{\varphi}_{mn \times 1} := \text{vec}\left(\varphi\right)$, $b_{\left(m+n\right) \times 1} := \left[\alpha_{m\times1}, \beta_{n\times1}\right]^{\top}$. If we let $e_{n} := [\underbrace{1, 1, \hdots, 1}_{n\,\text{times}}]^{\top}$, then the implementation $A_{\left(m+n\right)\times mn} =  \begin{bmatrix} e_{n}^{\top} \otimes I_{m}\\I_{n} \otimes e_{m}^{\top}\end{bmatrix}$ was found to achieve fast offline construction of the constraint matrix.

\subsubsection{Storage complexity}
For $m=n$, the constraint matrix $A$ in (\ref{StandardLPform}), is a binary matrix of size $2 n \times n^{2}$, whose each row has $n$ ones. Consequently, there are total $2 n^{2}$ ones in the constraint matrix and the remaining $2 n^{2}\left(n - 1\right)$ elements are zero. Hence at any fixed time, the sparse representation of the constraint matrix needs \# non-zero elements $\times 3 = 6 n^{2}$ storage. The PMF vectors are, in general, fully populated. In addition, we need to store the model and true sample coordinates, each of them being a $n_{o}$-tuple. Hence at any fixed time, constructing cost matrix requires storing $2 n_{o} n$ values. Thus total storage complexity at any given snapshot, is $2 n \left(3n + n_{o} + 1\right) = O\left(n^{2}\right)$, assuming $n>n_{o}$. However, if the sparsity of constraint matrix is not exploited by the solver, then storage complexity rises to $2n \left(n^{2} + n_{o} + 1\right) = O\left(n^{3}\right)$. For example, if we take $n = 1000$ samples and use double precision arithmetic, then solving the LP at each time requires either megabytes or gigabytes of storage, depending on whether or not sparse representation is utilized by the solver\footnote{We used MOSEK (available at \texttt{www.mosek.com}) as the LP solver.}. For $m \neq n$, it is easy to verify that the sparse storage complexity is $\left(6mn + \left(m+n\right)n_{o} + m + n\right)$, and the non-sparse storage complexity is $\left(m+n\right)\left(mn + n_{0} + 1\right)$.

%%%%%%%%%%%%%%%%%%%%%%%%%%%%%%%%%%%%%%%%%%%%%%%%%%%%%%%%%%%%%%%%%%%%%%%%%%%%%%%%%%%%%%%%%%%%%%%%%%%%%%%%%%%%%%%%%%%

\section{Construction of Validation Certificates}

\subsection{Probabilistically robust model validation}
Often in practice, the exact initial density is not known to facilitate our model validation framework; instead a class of densities may be known. For example, it may be known that the initial density is symmetric unimodal but its exact shape (e.g. normal, semi-circular etc.) may not be known. Even when the distribution-type is known (e.g. normal), it is often difficult to pinpoint the parameter values describing the initial density function. To account such scenarios, consider a random variable $\Delta : \Omega \rightarrow E$, that induces a probability triplet $\left(\Omega, \mathcal{F},\mathbb{P}\right)$ on the space of initial densities. Here $E \subset \Omega$ and $\#\left(E\right) = 1$. The random variable $\Delta$ picks up an initial density from the collection of admissible initial densities $\Omega := \{\xi_{0}^{(1)}\left(\widetilde{x}\right), \xi_{0}^{(2)}\left(\widetilde{x}\right), \hdots\}$ according to the law of $\Delta$. For example, if we know $\xi_{0} \sim \mathcal{N}\left(\mu_{0}, \sigma_{0}^{2}\right)$ with a given joint distribution over the $\left(\mu_{0}, \sigma_{0}^{2}\right)$ space, then in our model validation framework, one sample from this space will return one distance measure between the instantaneous output PDFs. How many such $\left(\mu_{0}, \sigma_{0}^{2}\right)$ samples are necessary to guarantee the robustness of the model validation oracle? The Chernoff bound provides such an estimate for finite sample complexity.

At time step $t_{k}$, let the \emph{validation probability} be $p\left(\gamma_{k}\right) := \mathbb{P}\left(_{2}W_{2}\left(\eta_{k}\left(y\right), \widehat{\eta}_{k}\left(\widehat{y}\right)\right) \leqslant \gamma_{k}\right)$. Here $\gamma_{k} \in \mathbb{R}^{+}$ is the prescribed instantaneous tolerance level. If the model validation is performed by drawing $N$ samples from $\Omega$, then the \emph{empirical validation probability} is $\widehat{p}_{N}\left(\gamma_{k}\right) := \displaystyle\frac{1}{N}\displaystyle\sum_{i=1}^{N} \chi_{V_{k}^{\left(i\right)}}$ where $V_{k}^{\left(i\right)} := \{\widehat{\eta}_{k}^{\left(i\right)}\left(\widehat{y}\right) \: : \: _{2}W_{2}\left(\eta_{k}^{\left(i\right)}\left(y\right), \widehat{\eta}_{k}^{\left(i\right)}\left(\widehat{y}\right)\right) \: \leqslant \: \gamma_{k}\}$. Consider $\epsilon, \delta$ $\in \left(0,1\right)$ as the desired accuracy and confidence, respectively.
\begin{lemma}
(\textbf{Chernoff bound})\cite{Tempo2004} For any $\epsilon, \delta$ $\in \left(0,1\right)$, if $N \geqslant N_{\emph{ch}} := \displaystyle\frac{1}{2\epsilon^{2}}\log\displaystyle\frac{2}{\delta}$, then $\mathbb{P}\left(\lvert p\left(\gamma_{k}\right) - \widehat{p}_{N}\left(\gamma_{k}\right) \lvert < \epsilon\right) > 1 - \delta$.
\label{ChernoffBound}
\end{lemma}
The above lemma allows us to construct \emph{probabilistically robust validation certificate} (PRVC) $\widehat{p}_{N}\left(\gamma_{k}\right)$ through the algorithm below.
\begin{algorithm}[t]
\caption{Construct PRVC}
\label{AlgoPRVC}
\begin{algorithmic}[1]
\footnotesize{\Require $\epsilon, \delta$ $\in \left(0,1\right)$, $T$, $\nu$, law of $\Delta$, experimental data $\{\eta_{k}\left(y\right)\}_{k=1}^{\tau}$, model, tolerance vector $\{\gamma_{k}\}_{k=1}^{\tau}$
\State $N \leftarrow N_{\text{ch}}\left(\epsilon, \delta\right)$ \Comment{Using lemma \ref{ChernoffBound}}
\State Draw random functions $\xi_{0}^{(1)}\left(\widetilde{x}\right), \xi_{0}^{(2)}\left(\widetilde{x}\right), \hdots, \xi_{0}^{(N)}\left(\widetilde{x}\right)$ according to the law of $\Delta$
\For {$k=1$ to $\tau$} \Comment{Index for time step}
\For {$i=1$ to $N$} \Comment{Index for initial density}
\For {$j=1$ to $\nu$}\Comment{Samples drawn from $\xi_{0}^{(i)}\left(\widetilde{x}\right)$}
\State Propagate states using dynamics
\State Propagate measurements
\EndFor
\State Propagate $\widehat{\xi}_{k}^{\left(i\right)}\left(\widehat{\widetilde{x}}\right)$ \Comment{Use (\ref{Liouville}), (\ref{FokkerPlanck}), (\ref{PFoperator}) or (\ref{StocMapPDFupdate})}
\State Compute $\widehat{\eta}_{k}^{\left(i\right)}\left(\widehat{y}\right)$
\State Compute $_{2}W_{2}\left(\eta_{k}^{\left(i\right)}\left(y\right), \widehat{\eta}_{k}^{\left(i\right)}\left(\widehat{y}\right)\right)$ \Comment{Distributional comparison by solving LP (\ref{HitchcockKoopmansLP}) subject to (C0)--(C3)}
\State sum $\leftarrow$ 0 \Comment{Initialize}
\If {$_{2}W_{2}\left(\eta_{k}^{\left(i\right)}\left(y\right), \widehat{\eta}_{k}^{\left(i\right)}\left(\widehat{y}\right)\right) \leqslant \gamma_{k}$}
\State sum $\leftarrow$ sum + 1
\EndIf
\EndFor
\State $\widehat{p}_{N}\left(\gamma_{k}\right) \leftarrow \displaystyle\frac{\text{sum}}{N}$ \Comment{Construct PRVC vector}
\EndFor}
\end{algorithmic}
\end{algorithm}
The PRVC vector, with $\epsilon$ accuracy, returns the probability that the model is valid at time $t_{k}$, in the sense that the instantaneous output PDFs are no distant than the required tolerance level $\gamma_{k}$. Lemma \ref{ChernoffBound} lets the user control the accuracy $\epsilon$ and the confidence $\delta$, with which the preceding statement can be made. Thus the framework enables us to compute a provably correct validation certificate on the face of uncertainty with finite sample complexity.

\subsection{Probabilistically worst-case model validation}
Following \cite{KhargonekarTikku1996,Tempo1997,ChenZhou1998}, one can also define a probabilistic notion of the worst-case model validation performance as $\gamma_{k}^{\text{wc}} := \underset{\Delta}{\sup}\, _{2}W_{2}\left(\eta_{k}\left(y\right), \widehat{\eta}_{k}\left(\widehat{y}\right)\right)$, and its empirical estimate $\widehat{\gamma}_{k}^{N} := \underset{i=1,\hdots,N}{\max}\: _{2}W_{2}\left(\eta_{k}^{\left(i\right)}\left(y\right), \widehat{\eta}_{k}^{\left(i\right)}\left(\widehat{y}\right)\right)$. The sample complexity for probabilistically worst-case model validation is given by the lemma below.
\begin{lemma}
(\textbf{Worst-case bound}) (p. 128, \cite{Tempo2004}) For any $\epsilon, \delta$ $\in \left(0,1\right)$, if $N \geqslant N_{\text{wc}} := \displaystyle\frac{\log\displaystyle\frac{1}{\delta}}{\log\displaystyle\frac{1}{1-\epsilon}}$, then $\mathbb{P}\left(\mathbb{P}\left( _{2}W_{2}\left(\eta_{k}\left(y\right), \widehat{\eta}_{k}\left(\widehat{y}\right)\right) \leqslant \widehat{\gamma}_{k}^{N}\right) \geqslant 1 - \epsilon\right) > 1 - \delta$.
\label{worstcasebound}
\end{lemma}
Notice that in general, there is no guarantee that the empirical estimate $\widehat{\gamma}_{k}^{N}$ is close to the true worst-case performance $\gamma_{k}^{\text{wc}}$. Also, the performance bound is obtained \emph{a posteriori} while the robust validation framework accounted for \emph{a priori} tolerance levels. The corresponding \emph{probabilistically worst-case validation certificate} (PWVC) $\widehat{\gamma}_{k}^{N}$ can be computed from the following algorithm.
\begin{algorithm}[t]
\caption{Construct PWVC}
\label{AlgoPWVC}
\begin{algorithmic}[1]
\small{\Require $\epsilon, \delta$ $\in \left(0,1\right)$, $\tau$, $\nu$, law of $\Delta$, experimental data $\{\eta_{k}\left(y\right)\}_{k=1}^{\tau}$, model
\State $N \leftarrow N_{\text{wc}}\left(\epsilon, \delta\right)$ \Comment{Using lemma \ref{worstcasebound}}
\State Draw $N$ random functions $\xi_{0}^{(1)}\left(\widetilde{x}\right), \xi_{0}^{(2)}\left(\widetilde{x}\right), \hdots, \xi_{0}^{(N)}\left(\widetilde{x}\right)$ according to the law of $\Delta$ \Comment{Use MCMC}
\For {$k=1$ to $\tau$} \Comment{Index for time step}
\For {$i=1$ to $N$} \Comment{Index for initial density}
\For {$j=1$ to $\nu$}\Comment{Index for samples in the extended state space, drawn from $\xi_{0}^{(i)}\left(\widetilde{x}\right)$}
\State Propagate states using dynamics
\State Propagate measurements
\EndFor
\State Propagate $\widehat{\xi}_{k}^{\left(i\right)}\left(\widehat{\widetilde{x}}\right)$ \Comment{Use (3), (7), (9) or (11)}
\State Compute $\widehat{\eta}_{k}^{\left(i\right)}\left(\widehat{y}\right)$ \Comment{Algebraic transformation}
\State Compute $_{2}W_{2}\left(\eta_{k}^{\left(i\right)}\left(y\right), \widehat{\eta}_{k}^{\left(i\right)}\left(\widehat{y}\right)\right)$ \Comment{Distributional comparison by solving LP}
\State $\widehat{\gamma}_{k}^{N} \leftarrow \underset{i=1,\hdots,N}{\max}\: _{2}W_{2}\left(\eta_{k}^{\left(i\right)}\left(y\right), \widehat{\eta}_{k}^{\left(i\right)}\left(\widehat{y}\right)\right)$ \Comment{Empirically estimate worst-case performance}
\EndFor
\EndFor}
\end{algorithmic}
\end{algorithm}
In summary, the algorithm, with high probability $\left(1-\epsilon\right)$, only ensures that the output PDFs are at most $\widehat{\gamma}_{k}^{N}$ far. The preceding statement can be made with probability at least $1-\delta$.

%%%%%%%%%%%%%%%%%%%%%%%%%%%%%%%%%%%%%%%%%%%%%%%%%%%%%%%%%%%%%%%%%%%%%%%%%%%%%%%%%%%%%%%%%%%%%%%%%%%%%%%%%%%%%%%%%%%

\section{Illustrative Examples}

\begin{figure}[th]
\centering
\includegraphics[width=2.7in]{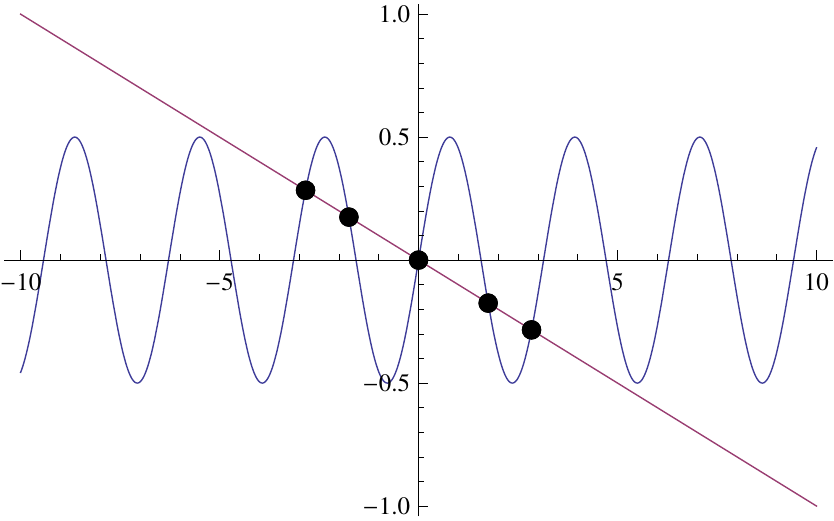}
\caption{\small{Points of intersection of the curve $f\left(x\right) = b \sin 2x$ and the line $g\left(x\right) = - a x$.}}
\label{FixedPtAbscissa}
\includegraphics[width=2.5in]{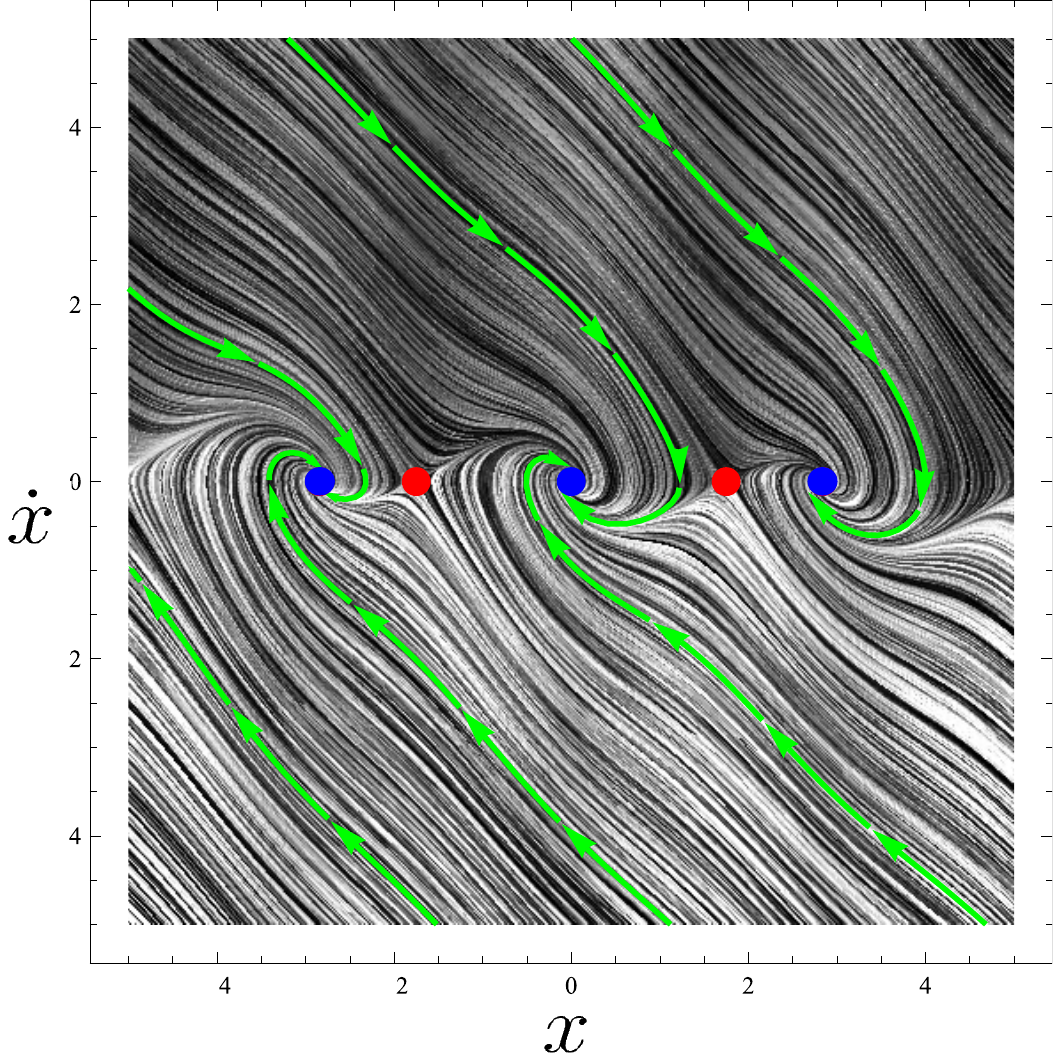}
\caption{\small{Phase portrait of the vector field (\ref{ContDeterministic}) with three stable and two saddle fixed points.}}
\label{ContDetPhasePortrait}
\end{figure}

\begin{example}Continuous-time deterministic dynamics\end{example}
Consider the following nonlinear dynamical system
\begin{eqnarray}
\ddot{x} = - a x - b \sin 2x - c \dot{x}, \quad a=0.1, b=0.5, c=1.
\label{ContDeterministic}
\end{eqnarray}
The system has five fixed points $P_{0} = \left(0,0\right)$, $P_{1}^{\pm} = \left(\pm 1.7495, 0\right)$, $P_{2}^{\pm} = \left(\pm 2.8396, 0\right)$, which can be solved by noting the abscissa values of the points of intersection of two curves $f\left(x\right) = b \sin 2x$ and $g\left(x\right) = - a x$, as shown in Fig \ref{FixedPtAbscissa}. From linear analysis, it is easy to verify that $P_{0}$ and $P_{2}^{\pm}$ are stable foci while $P_{1}^{\pm}$ are saddles (Fig. \ref{ContDetPhasePortrait}).
To illustrate our model validation framework, let's assume that `true data' is generated by the dynamics (\ref{ContDeterministic}). However, this true dynamics is unknown to the modeler, whose proposed model is a linearization of (\ref{ContDeterministic}) about the origin. We emphasize here that the purpose of (\ref{ContDeterministic}) is only to create the synthetic data and to demonstrate the proof-of-concept. In a realistic model validation, the data arrives from experimental measurements, not from another model. For simplicity, we take the outputs same as states for both true and model dynamics.

\begin{figure*}[thb]
\begin{minipage}[b]{0.5\linewidth}
\centering
%\begin{lpic}[l(),r(),t(),b(),draft,clean]{NormalizedWassNoMarking(0.35)}
% \lbl[r]{65,56.3; \footnotesize{$_{2}W_{2}$}
% \lbl[r]{60,6.2; \footnotesize{Pathfinder (1997)}}
% \end{lpic}
\hspace*{-0.2in}
\includegraphics[width=3.1in]{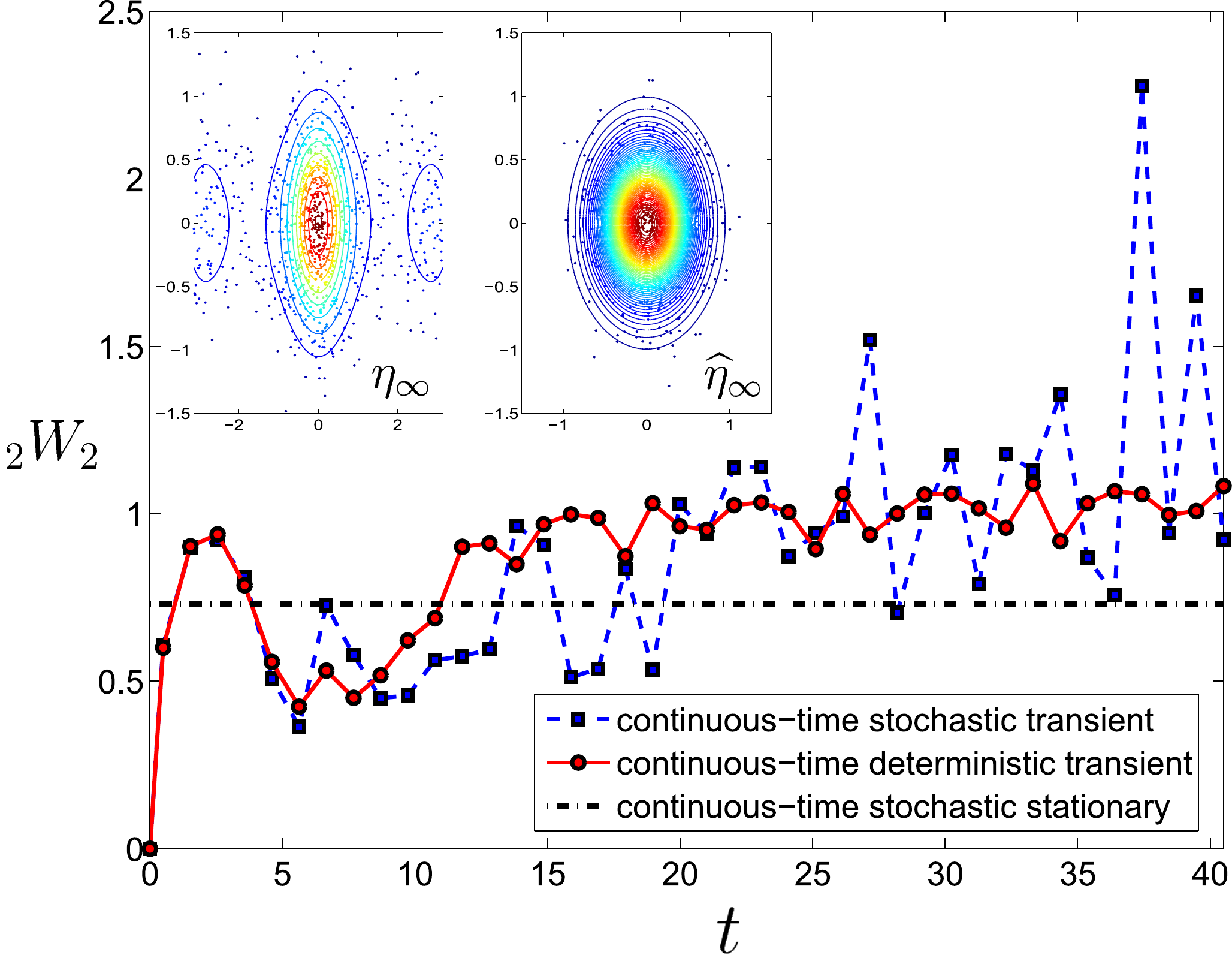}
\vspace*{-0.1in}
\caption{\small{Starting with $\xi_{0} = \mathcal{U}\left(\left[-\pi, \pi\right] \times \left[-\pi, \pi\right]\right)$, the \emph{solid line} shows time history of $_{2}W_{2}$ measured between the joint state PDFs for (\ref{ContDeterministic}) and its linearization about the origin. The \emph{dashed line} shows the same between (\ref{ContStochastic}) and its linearization about the origin. The \emph{dash-dotted line} shows the stationary $_{2}W_{2}$ between known $\eta_{\infty}$ and $\widehat{\eta}_{\infty}$ (\emph{contours in the inset plot}), given by (\ref{StationaryDensityContStoc}) and (\ref{LinearStationaryDensity}) respectively, and is computed by solving the optimal transport LP between their MCMC samples (\emph{scattered points in the inset plot}). All computations were done with 1000 Halton samples \cite{Niederreiter1992} from $\xi_{0}$ and 50 eigenfunctions in noise KL expansion.}}
\label{Wcontinuous}
\end{minipage}
\hspace{0.1cm}
\begin{minipage}[b]{0.5\linewidth}
\vspace*{-5in}
\centering
\vspace*{-5in}
\includegraphics[width=3in]{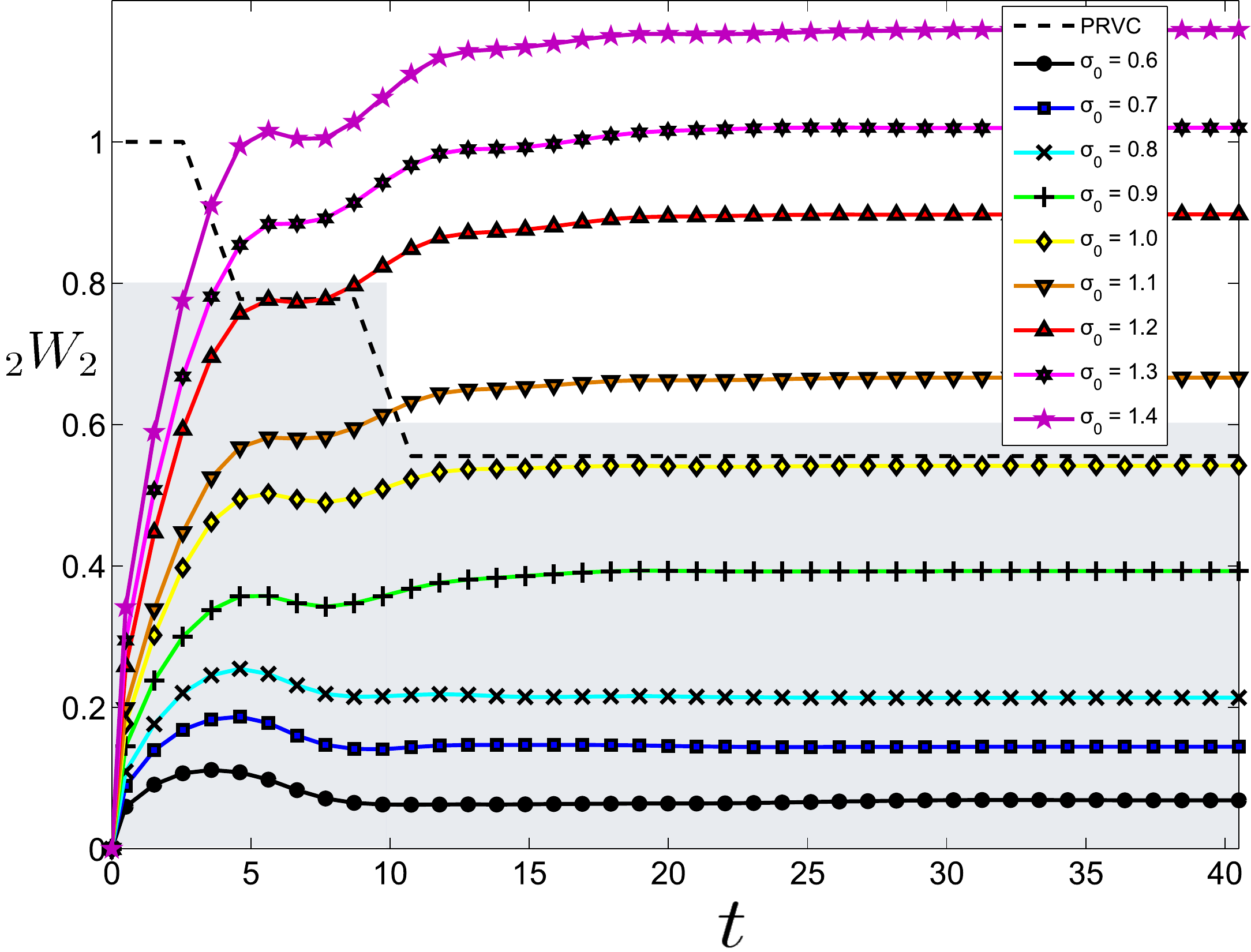}

\caption{\small{Starting with $\xi_{0} = \mathcal{N}\left(0, \sigma_{0}^{2}I_{2}\right)$, transient Wasserstein time histories, measured between the joint state PDFs for (\ref{ContDeterministic}) and its linearization about the origin. In this case, increasing $\sigma_{0}$ increases $_{2}W_{2}$ \emph{at all times}. Further, notice that $_{2}W_{2}$ trajectories with larger $\sigma_{0}$, converges to higher asymptotic values. This is due to the fact that the stationary density of (\ref{ContDeterministic}) is of the form $\eta_{\infty}(y) = \sum_{i=1}^{5} m_{i}^{\star} \delta\left(y - y_{i}^{\star}\right)$, and hence depends on $\xi_{0}$, as explained in Remark \ref{MassFractionRemark} and fourth row of Table \ref{WassTable}. The \emph{shaded area} shows user-specified tolerance level $\{\gamma_{j}\}_{j=1}^{40}$, from which PRVC (\emph{dashed line}) can be computed. In this case, PWVC is simply the $_{2}W_{2}$ trajectory corresponding to $\sigma_{0} = 1.4$.}}
\label{WDiscrimination}
\end{minipage}
\end{figure*}

Starting from the bivariate uniform distribution $\mathcal{U}\left(\left[-\pi, \pi\right] \times \left[-\pi, \pi\right]\right) =: \xi_{0}$, we evolve the respective joint PDFs $\xi = \eta$ and $\widehat{\xi} = \widehat{\eta}$, through true and model dynamics via MOC implementation of Liouville equation \cite{HalderBhattacharya2011}. The distributional shape discrepancy is captured via the Wasserstein gap $\left( _{2}W_{2}\left(\eta, \widehat{\eta}\right)\right)$ between these instantaneous joint PDFs, as shown in Fig. \ref{Wcontinuous} (\emph{solid line}), computed by solving the LP (\ref{HitchcockKoopmansLP}), (C1)--(C3). As the individual joint PDFs converge toward their respective stationary densities, the slope of the Wasserstein time-history decreases progressively. Fig. \ref{WDiscrimination} shows the Wasserstein gap trajectories when $\xi_{0}$ is taken to be $\mathcal{N}\left(0,\sigma_{0}^{2}I_{2}\right)$, instead of uniform. In this case, we observe that larger initial dispersion causes larger Wasserstein gap. Suppose the user-specified tolerance level $\{\gamma_{j}\}_{j=1}^{40}$ is $0.8$ for first 10 instances and $0.6$ for next 30 instances of measurement availability, as shown by the shaded area in Fig. \ref{WDiscrimination}. Given the set of admissible initial densities $\{\xi_{0}^{(1)}, \hdots, \xi_{0}^{(9)}\}$ with $\xi_{0}^{(i)} := \mathcal{N}\left(0,\sigma_{0i}^{2}I_{2}\right)$, $i=1,\hdots,9$, we can compute the PRVC vector, shown as the dashed line in Fig. \ref{WDiscrimination}, to be $\left[\underbrace{1, \hdots, 1,}_{3 \:\text{times}} 0.89, \underbrace{0.78, \hdots, 0.78,}_{5\:\text{times}} 0.67, \underbrace{0.56, \hdots, 0.56}_{30 \:\text{times}}\right]^{\top}$.

\begin{example}Continuous-time stochastic dynamics\end{example}
 Here we assume the true data to be generated by (\ref{ContDeterministic}) with additive white noise having autocorrelation $Q \delta\left(t_{1} - t_{2}\right)$, $t_{1}, t_{2} \geqslant 0$. Letting $x_{1} = x$ and $x_{2} = \dot{x}$, the associated It$\hat{\text{o}}$ SDE can be written in state-space form similar to (\ref{ItoSDE})
\begin{eqnarray}
\begin{Bmatrix}
dx_{1}\\dx_{2}
\end{Bmatrix} =
\begin{Bmatrix}
x_{2}\\- a x_{1} - b \sin 2x_{1} - c x_{2}
\end{Bmatrix} \: dt +
\begin{Bmatrix}
0\\1
\end{Bmatrix} \: d\beta,
\label{ContStochastic}
\end{eqnarray}
where $\beta\left(t\right)$ is a Wiener process with autocorrelation $Q \left(t_{1} \wedge t_{2}\right)$. The stationary Fokker-Planck equation for (\ref{ContStochastic}) can be solved in closed form (Appendix E)
\begin{eqnarray}
\eta_{\infty}\left(x_{1}, x_{2}\right) \propto \exp\left(- \displaystyle\frac{c}{2Q} \left(a x_{1}^{2} + x_{2}^{2} - b \cos 2x_{1}\right)\right),
\label{StationaryDensityContStoc}
\end{eqnarray}
and one can verify that peaks of (\ref{StationaryDensityContStoc}) appear at the fixed points of the nonlinear drift.

Let the proposed model be the linearization of (\ref{ContStochastic}) about the origin. It is well-known \cite{LiberzonBrockett2000} that the stationary density of a linear SDE of the form $d\widehat{x} = A \widehat{x} \: dt + B \: d\beta$, is given by
\begin{eqnarray}
\widehat{\eta}_{\infty}\left(\widehat{x}\right) = \mathcal{N}\left(\mathbf{0}, \Sigma_{\infty}\right) = \displaystyle\frac{\exp\left({-\frac{1}{2} \:\: \widehat{x}^{\top} \Sigma_{\infty}^{-1} \widehat{x}}\right)}{\sqrt{\left(2\pi\right)^{2} \text{det}\left(\Sigma_{\infty}\right)}} ,
\label{LinearStationaryDensity}
\end{eqnarray}
provided $A$ is Hurwitz and $\left(A, B\right)$ is a controllable pair. The steady-state covariance matrix $\Sigma_{\infty}$ solves $A \Sigma_{\infty} + \Sigma_{\infty} A^{\top} + B Q B^{\top} = 0$. For the linearized version of (\ref{ContStochastic}), $A = \begin{bmatrix} 0 & 1\\ \left(-a - 2b\right) & -c \end{bmatrix}$ and $B = \begin{Bmatrix} 0 \\ 1 \end{Bmatrix}$ satisfy the aforementioned conditions and the stationary density is obtained from (\ref{LinearStationaryDensity}).

Taking the initial density same as in Example 3.1, we propagated the joint PDFs for (\ref{ContStochastic}) and the linear SDE using the KLPF method described in \cite{Dutta2011}. The \emph{dashed line} in Fig. \ref{Wcontinuous} shows the Wasserstein trajectory for this case. The dash-dotted line in Fig. \ref{Wcontinuous} shows the asymptotic Wasserstein gap between the respective stationary densities (\ref{StationaryDensityContStoc}) and (\ref{LinearStationaryDensity}). Due to randomized sampling, all stochastic computations are in probabilistically approximate sense \cite{vidyasagar2001randomized}.

\begin{figure*}[t]
\centering
\vspace*{0.1in}
\includegraphics[width=7in]{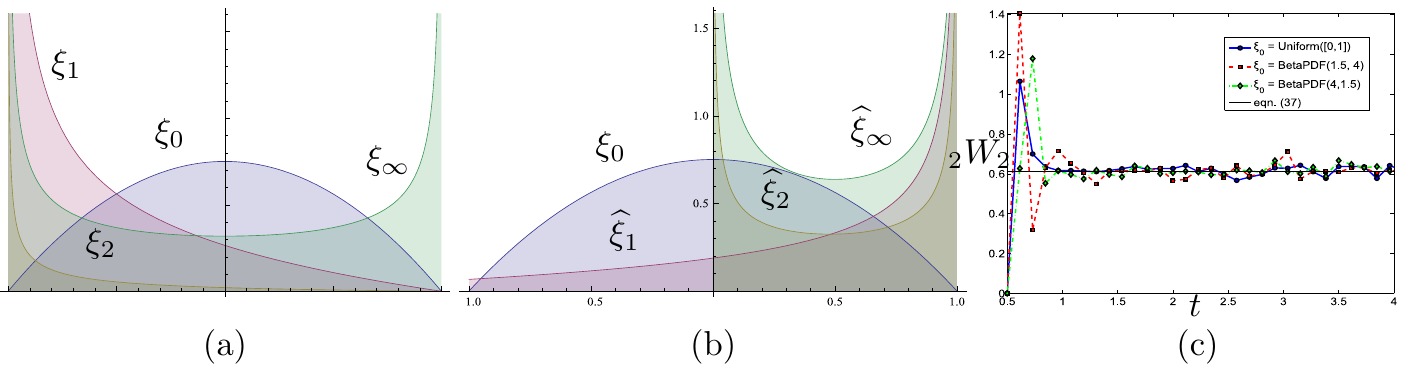}
\caption{\small{Starting with $\xi_{0}\left(x\right) = \frac{3}{4}\left(1-x^{2}\right)$, evolution of PDFs for (a) \emph{true} PF operator (\ref{PFoperatorChebyshev}), and (b) \emph{model} PF operator (\ref{PFoperatorLogistic}). (c) Wasserstein time histories between PF operators (\ref{PFoperatorChebyshev}) and (\ref{PFoperatorLogistic}) for various initial PDFs.}}
\label{MixedBagExamples}
\end{figure*}

\begin{example}Discrete-time deterministic dynamics\end{example}
Let the true data be generated by the Chebyshev map \cite{GeiselFairen1984} $\mathcal{T} : \left[-1, 1\right] \mapsto \left[-1, 1\right]$, given by
\begin{eqnarray}
x_{k+1} = \mathcal{T}\left(x_{k}\right) = \cos\left(2 \cos^{-1} x_{k}\right).
\label{ChebyshevMap}
\end{eqnarray}
If we let $\xi_{k} := \xi\left(x_{k}\right)$, then the PF operator $\mathscr{P}: \xi_{k} \mapsto \xi_{k+1}$, for (\ref{ChebyshevMap}) can be computed \cite{MackeyTyranKaminska2006} as
\begin{eqnarray}
\mathscr{P} \xi_{k} = \displaystyle\frac{1}{2 \sqrt{2x_{k} + 2}} \left[\xi\left(\sqrt{\displaystyle\frac{x_{k}+1}{2}}\right) + \xi\left(- \sqrt{\displaystyle\frac{x_{k}+1}{2}}\right)\right],
\label{PFoperatorChebyshev}
\end{eqnarray}
\noindent with stationary PDF $\xi_{\infty}\left(x\right) = \displaystyle\frac{1}{\pi \sqrt{1 - x^{2}}}$, and CDF $F_{\infty}\left(x\right) = \displaystyle\frac{2}{\pi} \sin^{-1}\left(\sqrt{\displaystyle\frac{x+1}{2}}\right)$. Notice that for small $x_{k}$, (\ref{ChebyshevMap}) behaves like a quadratic transformation. Suppose the following logistic map $\widehat{\mathcal{T}} : \left[0, 1\right] \mapsto \left[0, 1\right]$, is proposed to model the data generated by (\ref{ChebyshevMap}):
\begin{eqnarray}
x_{k+1} = \widehat{\mathcal{T}}\left(\widehat{x}_{k}\right) = 4 \widehat{x}_{k}\left(1 - \widehat{x}_{k}\right).
\label{LogisticMap}
\end{eqnarray}
The PF operator $\widehat{\mathscr{P}}: \widehat{\xi}_{k} \mapsto \widehat{\xi}_{k+1}$, for (\ref{LogisticMap}) is given by \cite{LasotaMackey1994}
\begin{eqnarray}
\widehat{\mathscr{P}} \widehat{\xi}_{k} = \displaystyle\frac{1}{4 \sqrt{1 - \widehat{x}_{k}}} \left[\widehat{\xi}\left(\displaystyle\frac{1 + \sqrt{1 - \widehat{x}_{k}}}{2}\right) + \widehat{\xi}\left(\displaystyle\frac{1 - \sqrt{1 - \widehat{x}_{k}}}{2}\right)\right],
\label{PFoperatorLogistic}
\end{eqnarray}
with stationary PDF $\widehat{\xi}_{\infty}\left(\widehat{x}\right) = \displaystyle\frac{1}{\pi\sqrt{\widehat{x} \left(1 - \widehat{x}\right)}}$, and CDF $\widehat{F}_{\infty}\left(\widehat{x}\right) = \displaystyle\frac{2}{\pi} \sin^{-1}\left(\sqrt{\widehat{x}}\right)$. Taking the outputs identical to states, the asymptotic Wasserstein distance between (\ref{ChebyshevMap}) and (\ref{LogisticMap}), becomes
\begin{eqnarray}
&& _{2}W_{2}\left(\xi_{\infty}\left(x\right), \widehat{\xi}_{\infty}\left(\widehat{x}\right)\right) = \sqrt{\displaystyle\int_{0}^{1} \left(F_{\infty}^{-1}\left(\varsigma\right) - \widehat{F}_{\infty}^{-1}\left(\varsigma\right)\right)^{2} \: d\varsigma} \nonumber\\
&=& \sqrt{\displaystyle\int_{0}^{1} \left(2\sin^{2}\left(\frac{\pi\varsigma}{2}\right) - 1 - \sin^{2}\left(\frac{\pi\varsigma}{2}\right)\right)^{2} \: d\varsigma} \nonumber\\
&=& \sqrt{\displaystyle\int_{0}^{1} \left(\frac{1}{2} + \frac{\cos\left(\pi\varsigma\right)}{2}\right)^{2} \: d\varsigma} \approx 0.6124.
\label{AsympWassChebyshevLogistic}
\end{eqnarray}
Given an initial density $\xi_{0}$, the transient PDFs $\xi\left(x,t\right)$ and $\widehat{\xi}\left(x,t\right)$ can be computed from (\ref{PFoperatorChebyshev}) and (\ref{PFoperatorLogistic}) (Fig. \ref{MixedBagExamples}(a) and (b)). Fig. \ref{MixedBagExamples}(c) shows the transient Wasserstein time-histories $_{2}W_{2}\left(\xi\left(x,t\right),\widehat{\xi}\left(x,t\right)\right)$ for various initial PDFs, which converge to its asymptotic value obtained analytically in (\ref{AsympWassChebyshevLogistic}).

\begin{example}Discrete-time stochastic dynamics\end{example}

Consider the true data being generated from the logistic map with multiplicative stochastic perturbation:
\begin{eqnarray}
x_{k+1} = \mathcal{T}\left(x_{k}, \zeta_{k}\right) = \zeta_{k}\mathcal{S}\left(x_{k}\right) = \zeta_{k} x_{k}\left(1 - x_{k}\right),
\label{LogisticMultiplicativeNoise}
\end{eqnarray}
where $\mathcal{S} : \left[0, 1\right] \mapsto \left[0, 1\right]$, and $\{\zeta_{k}\}_{0}^{\infty}$ are i.i.d random variables on $\left[0, 4\right]$, drawn from noise density $\phi(.)$. This map has found applications in population dynamics and size-dependent branching processes \cite{AthreyaDai2000,Klebaner1997}. The PF operator for (\ref{LogisticMultiplicativeNoise}) is given by (p. 330-331, \cite{LasotaMackey1994})
\begin{eqnarray}
\mathscr{P}\xi_{k} = \displaystyle\int_{0}^{\infty} \xi\left(y\right) \mathcal{K}_{\text{mul}}\left(x_{k}, y\right) \: dy,
\label{PFoperatorMultiplicativeStochastic}
\end{eqnarray}
with the \emph{multiplicative} stochastic kernel $\mathcal{K}_{\text{mul}}\left(x_{k}, y\right) := \displaystyle\frac{1}{\mathcal{S}\left(y\right)} \:\phi\left(\displaystyle\frac{x_{k}}{\mathcal{S}\left(y\right)}\right)$. In particular, $\zeta_{k} \sim \mathcal{N}\left(0,1\right)$ results $\mathscr{P}\xi_{k} = \displaystyle\int_{0}^{\infty} \xi\left(y\right) \displaystyle\frac{1}{\sqrt{2\pi} \: y\left(1 - y\right)} \: e^{- \frac{1}{2}\frac{x^{2}}{y^{2}\left(1 - y\right)^{2}}} \, dy$. The asymptotic behavior of (\ref{LogisticMultiplicativeNoise}) is known \cite{AthreyaDai2000} to depend on the noise density $\phi\left(.\right)$. Specifically, $\mathbb{E}\left[\log \zeta_{0}\right] < 0, = 0$, and $> 0$ implies $x_{k}\ \xrightarrow{\text{a.s.}}\ 0$, $x_{k}\ \xrightarrow{\text{i.p.}}\ 0$, and existence of stationary density $\xi_{\infty}$ on $\left(0,1\right) \, \forall x_{0} \neq 0$, respectively. For example, if $\zeta_{k} \sim \mathcal{N}\left(0,1\right)$, then $\displaystyle\int_{0}^{4} \log\zeta \displaystyle\frac{1}{\sqrt{2 \pi}} e^{-\frac{\zeta^{2}}{2}} \: d\zeta = \text{erf}\left(2\sqrt{2}\right)\log\left(2\right) - 2 \sqrt{\displaystyle\frac{2}{\pi}} \: _{2}F_{2}\left(\displaystyle\frac{1}{2},\displaystyle\frac{1}{2}; \displaystyle\frac{3}{2},\displaystyle\frac{3}{2}; -8\right) \approx -0.32 < 0$, and hence $x_{k}\ \xrightarrow{\text{a.s.}}\ 0$.

Let the proposed model be
\begin{eqnarray}
\widehat{x}_{k+1} = \widehat{\mathcal{T}}\left(\widehat{x}_{k}, \widehat{\zeta}_{k}\right) = \widehat{\mathcal{S}}\left(x_{k}\right) + \widehat{\zeta}_{k} = \widehat{x}_{k} + \widehat{\zeta}_{k},
\label{LinearMapAdditiveNoise}
\end{eqnarray}
where $\widehat{\mathcal{S}} : \mathbb{R} \mapsto \mathbb{R}$, and $\widehat{\zeta}_{k} \sim \mathcal{N}\left(0,1\right)$. The PF operator for a map with additive noise is of the form
\begin{eqnarray}
\widehat{\mathscr{P}}\widehat{\xi}_{k} = \displaystyle\int_{-\infty}^{\infty} \widehat{\xi}\left(y\right) \mathcal{K}_{\text{add}}\left(\widehat{x}_{k}, y\right) \: dy,
\label{PFoperatorLinearAdditiveNoise}
\end{eqnarray}
with the \emph{additive} stochastic kernel $\mathcal{K}_{\text{add}} \left(\widehat{x}_{k}, y\right) := \phi\left(\widehat{x}_{k} - \widehat{\mathcal{S}}\left(y\right)\right)$. Consequently, the PF operator for (\ref{LinearMapAdditiveNoise}) is $\widehat{\mathscr{P}}\widehat{\xi}_{k} = \displaystyle\int_{-\infty}^{\infty} \displaystyle\frac{1}{\sqrt{2\pi}} \exp\left(-\displaystyle\frac{\left(\widehat{x}_{k} - y\right)^{2}}{2}\right) \widehat{\xi}\left(y\right) \: dy$. It can be verified (p. 325, \cite{LasotaMackey1994}) that the successive iterate $\widehat{\mathscr{P}}^{k} \widehat{\xi}$ converges uniformly to zero as $k \rightarrow \infty$, and hence there is no non-trivial stationary density. Given an initial density, the Wasserstein distance can be computed between (\ref{PFoperatorMultiplicativeStochastic}) and (\ref{PFoperatorLinearAdditiveNoise}). This example demonstrates that (in)validating a stochastic map has sensitive dependence on noise density.

\begin{figure}[t]
\centering
\includegraphics[width=3in]{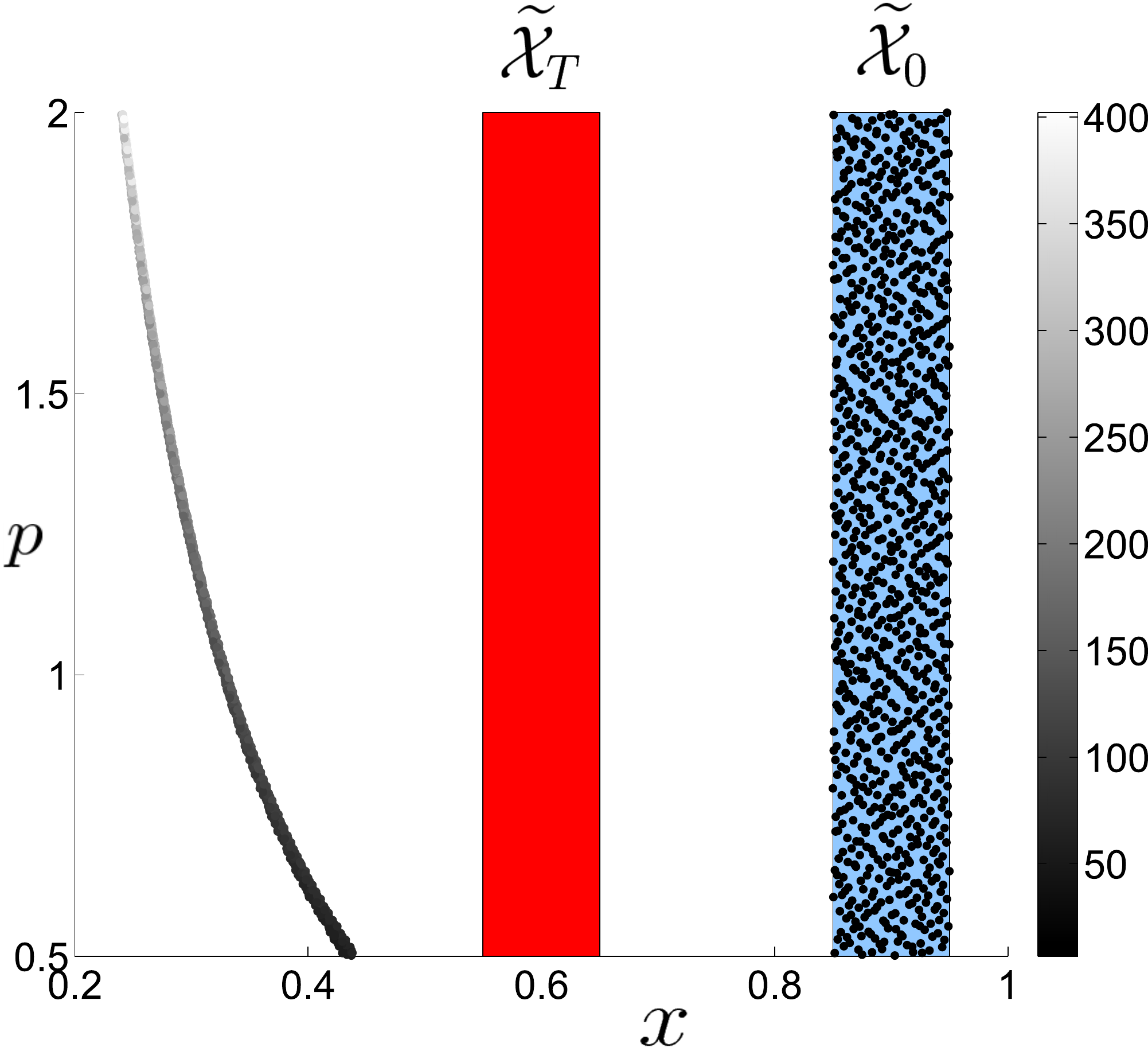}
\caption{\small{This plot illustrates how Prajna's barrier certificate-based invalidation
result can be recovered in our probabilistic
model validation framework. To show $\widetilde{\mathcal{X}}_{T}$ is not reachable
from the set $\widetilde{\mathcal{X}}_{0}$ in time T = 4, we sample $\widetilde{\mathcal{X}}_{0}$ uniformly
and propagate that uniform ensemble subject to the proposed model dynamics till
T = 4. The samples are gray-scale color coded (white = high probability, black = low
probability) according to the value of the joint PDF at that location. Here,
the model is invalidated since the pair of joint PDFs at initial and final time, does not satisfy the Liouville transport PDE corresponding to the model dynamics, as proved in Theorem \ref{RecoverPrajna}.}}
\label{DiscreteDetWass}
\end{figure}

\begin{example}Comparison with barrier certificate based model falsification\end{example}
Consider the nonlinear model validation problem stated as Example 4 in \cite{Prajna2006}, where the model is $\dot{x} = - p x^{3}$, with parameter $p \in \mathcal{P} = \left[0.5, 2\right]$. The measurement data are interval-valued sets $\mathcal{X}_{0} = \left[0.85, 0.95\right]$ at $t = 0$, and $\mathcal{X}_{T} = \left[0.55, 0.65\right]$ at $t = T = 4$. A barrier certificate of the form $B\left(x, t\right) = B_{1}\left(x\right) + t B_{2}\left(x\right)$ was found in \cite{Prajna2006} through sum-of-squares (SOS) optimization \cite{Prajna2002} where $B_{1}\left(x\right) = 8.35 x + 10.40 x^{2} - 21.50 x^{3} + 9.86 x^{4}$, and $B_{2}\left(x\right) = -1.78 + 6.58 x - 4.12 x^{2} - 1.19 x^{3} + 1.54 x^{4}$. The model was thereby invalidated by the existence of such certificate, i.e. the model $\dot{x} = - p x^{3}$, with parameter $p \in \mathcal{P}$ was shown to be inconsistent with measurements $\{\mathcal{X}_{0}, \mathcal{X}_{T}, T\}$.

To tackle this problem in our model validation framework, consider the spatio-temporal evolution of the joint PDF $\xi\left(x, p, t\right)$ over the extended state space $\widetilde{x} = \left[x \;\; p\right]^{\top}$, with initial support $\widetilde{\mathcal{X}}_{0} := \mathcal{X}_{0} \times \mathcal{P}$, under the action of the extended vector field $\widetilde{f}\left(x,p\right) = \left[-px^{3} \quad 0\right]^{\top}$. Our objective then, is to prove that for $T=4$, the PDF $\xi_{T}\left(x_{T}, p, T\right) = \mathcal{U}\left(x_{T}, p\right) = 1/\text{vol}\left(\widetilde{\mathcal{X}}_{T}\right)$ is not finite-time reachable from $\xi_{0}\left(x_{0}, p\right) = \mathcal{U}\left(x_{0}, p\right) = 1/\text{vol}\left(\widetilde{\mathcal{X}}_{0}\right)$, subject to the proposed model dynamics on the extended state space.

\begin{theorem}
The two-point boundary value problem
\begin{eqnarray}
\frac{\partial \xi}{\partial t} + \nabla_{\widetilde{x}} \cdot \left(\widetilde{f}\left(x,p\right) \xi\right) = \frac{\partial \xi}{\partial t} + \nabla_{x} \cdot \left(-px^{3} \xi\right) = 0, \nonumber\\
\xi\left(x(0), p, 0\right) = \xi_{0}\left(x_{0}, p\right) = \mathcal{U}\left(x_{0}, p\right) = 1/\text{\emph{vol}}\left(\widetilde{\mathcal{X}}_{0}\right), \nonumber\\
\xi\left(x(T), p, T\right) = \xi_{T}\left(x_{T}, p, T\right) = \mathcal{U}\left(x_{T}, p\right) = 1/\text{\emph{vol}}\left(\widetilde{\mathcal{X}}_{T}\right), \nonumber
\label{TPBVP}
\end{eqnarray}
has no solution for $\xi\left(x,p,t\right)$, such that $\displaystyle\int_{\widetilde{\mathcal{X}}(t)} \xi\left(x,p,t\right) \: dx \: dp = 1$, $\forall t \in \left(0,T\right)$.
\label{RecoverPrajna}
\end{theorem}

\begin{pf*}{Proof.}
MOC ODE \cite{HalderBhattacharya2011} corresponding to the Liouville PDE $\frac{\partial \xi}{\partial t} + \nabla_{\widetilde{x}} \cdot \left(\widetilde{f}\left(x,p\right) \xi\right) = 0$, yields a solution of the form
\begin{eqnarray}
\xi\left(x, p, t\right) = \xi_{0}\left(x_{0}, p\right) \exp\left(- \displaystyle\int_{0}^{t} \nabla_{\widetilde{x}} \cdot \left(\widetilde{f}\left(x\left(\tau\right), p\right)\right) d\tau \right).
\label{LiouvilleMOCgeneralFormSolution}
\end{eqnarray}
For the model dynamics $\dot{x} = - p x^{3}$, we have $\nabla_{\widetilde{x}} \cdot\left(\widetilde{f}\left(x\left(\tau\right),p\right)\right) = - 3 p \left(x\left(\tau\right)\right)^{2}$ and $\displaystyle\frac{1}{x^{2}} = \displaystyle\frac{1}{x_{0}^{2}} + 2 p t$. Consequently (\ref{LiouvilleMOCgeneralFormSolution}) results
\begin{eqnarray}
\xi\left(x, p, t\right) &=& \xi_{0}\left(x_{0}, p\right) \left(1 + 2 x_{0}^{2} p t\right)^{3/2} \nonumber\\
&=& \displaystyle\frac{1}{\left(1 - 2 x^{2} p t\right)^{3/2}} \, \xi_{0}\left(\pm\displaystyle\frac{x}{\sqrt{1 - 2 x^{2} p t}}, p\right).
\label{PrajnaCubicModelPDF}
\end{eqnarray}
In particular, for $\xi_{0}\left(x_{0}, p\right) = 1/\text{vol}\left(\widetilde{\mathcal{X}}_{0}\right)$, $\xi_{T}\left(x_{T}, p, T\right) = 1/\text{vol}\left(\widetilde{\mathcal{X}}_{T}\right)$, and $T = 4$, (\ref{PrajnaCubicModelPDF}) requires us to satisfy
\begin{eqnarray}
\left(1 - 8 x_{T}^{2} p\right) = \left(\displaystyle\frac{\text{vol}\left(\widetilde{X}_{T}\right)}{\text{vol}\left(\widetilde{X}_{0}\right)}\right)^{2/3} > 0 \Rightarrow
1 > 8 x_{T}^{2} p.
\label{ImpossibilityPrajnaCubic}
\end{eqnarray}
Since $8 x_{T}^{2} p$ is an increasing function in both $x_{T} \in \mathcal{X}_{T}$ and $p \in \mathcal{P}$, we need at least $1 > 8 \left(x_{T}\right)_{\text{min}}^{2} p_{\text{min}} = 8 \times \left(0.55\right)^{2} \times 0.5 = 1.21$, which is incorrect. Thus the PDF $\xi_{T}\left(x_{T}, p, T\right) \sim \mathcal{U}\left(x_{T}, p\right)$ is not finite-time reachable from $\xi_{0}\left(x_{0}, p\right) \sim \mathcal{U}\left(x_{0}, p\right)$ for $T = 4$, via the proposed model dynamics. Hence our measure-theoretic formulation recovers Prajna's invalidation result \cite{Prajna2006} as a special case. \hfill\qed
\end{pf*}

\begin{remark}{\textbf{(Relaxation of set-based invalidation)}}
Instead of binary (in)validation oracle, we can now measure the ``degree of validation" by computing the Wasserstein distance $_{2}W_{2}\left(\frac{1}{\left(1 - 2 x_{T}^{2} p T\right)^{3/2}} \, \frac{1}{\text{vol}\left(\widetilde{\mathcal{X}}_{0}\right)}, \; \frac{1}{\text{vol}\left(\widetilde{\mathcal{X}}_{T}\right)}\right)$ between the model predicted and experimentally measured joint PDFs. More importantly, it dispenses off the conservatism in barrier certificate based model validation by showing that the goodness of a model depends on the measures over the same pair of supports $\widetilde{\mathcal{X}}_{0}$ and $\widetilde{\mathcal{X}}_{T}$, than on the supports themselves. Indeed, given a joint PDF $\xi\left(x_{T}, p, T\right)$ supported over $\widetilde{\mathcal{X}}_{T}$ at $T=4$, from (\ref{PrajnaCubicModelPDF}) we can explicitly compute the initial PDF $\xi_{0}\left(x_{0}, p\right)$ supported over $\widetilde{\mathcal{X}}_{0}$ that, under the proposed model dynamics, yields the prescribed PDF, i.e.
\begin{eqnarray}
\xi_{0}\left(x_{0}, p\right) = \displaystyle\frac{1}{\left(1 + 8 x_{0}^{2} p\right)^{3/2}} \; \xi\left(\pm \displaystyle\frac{x_{0}}{\sqrt{1 + 8 x_{0}^{2} p}}, p, 4\right).
\label{KindOfInverseProblem}
\end{eqnarray}
In other words, if the measurements find the initial density given by (\ref{KindOfInverseProblem}) and final density $\xi\left(x_{T}, p, T\right)$ at $T = 4$, then the Wasserstein distance at $T = 4$ will be zero, thereby perfectly validating the model. This reinstates the importance of considering the \textbf{reachability of densities} over sets than \textbf{reachability of sets}, for model validation.
\end{remark}
\begin{remark}{\textbf{(Connections with Rantzer's density function-based invalidation)}}
Similar to barrier certificates, Rantzer's density functions \cite{Rantzer2001} can provide deductive invalidation guarantees (cf. Theorem 1 in \cite{RantzerPrajna2004}) by constructing a scalar function via convex program. Various applications of these two approaches for temporal verification problems have been reported \cite{PrajnaRantzer2007}. It is interesting to note that the main idea of Rantzer's density function stems from an integral form of Liouville equation, given by (cf. Lemma A.1 in \cite{Rantzer2001})
\begin{eqnarray}
\int_{\mathcal{X}_{T}} \xi \: dx \: - \: \int_{\mathcal{X}_{0}} \xi \: dx = \int_{0}^{T} \int_{\phi_{t}\left(\mathcal{X}_{0}\right)} \nabla_{x} \cdot \left(\xi f\right) dx \: dt,
\label{LiouvilleIntegralForm}
\end{eqnarray}
where the initial set $\mathcal{X}_{0}$ gets mapped to the set $\mathcal{X}_{T}$ at time $t=T$, under the action of the flow $\phi_{t}(\cdot)$ associated with the nonlinear dynamics $\dot{x} = f\left(x\right)$. The convex relaxation proposed for invalidation/safety verification (Theorem 1 in \cite{RantzerPrajna2004}), strives to construct an artificial ``density" $\xi = \xi_{r}\left(x,t\right)$ satisfying three conditions, viz. (i) $\xi_{r}\left(x,0\right) > 0, \: \forall \: x \in \mathcal{X}_{0}$, (ii) $\xi_{r}\left(x,T\right) \leqslant 0, \: \forall \: x \in \mathcal{X}_{T}$, and (iii) $\nabla_{x} \cdot \left(\xi_{r} f\right) \geqslant 0, \: \forall x \in \phi_{t}\left(\mathcal{X}_{0}\right), t \in \left(0, T\right)$. From (\ref{LiouvilleIntegralForm}), such a construction results a ``sign-based invalidation", and is only \textbf{sufficient} unless a Slater-like condition \cite{PrajnaRantzer2005} is satisfied. On the other hand, the ``validation in probability" framework proposed in this paper, relies on Liouville PDE-based exact arithmetic computation of $\xi$, and is a direct simulation-based non-deductive formulation. In this approach, model invalidation equals violation of (\ref{LiouvilleIntegralForm}), not just the sign-mismatch of its left-hand and right-hand side, and hence is \textbf{necessary and sufficient}. As shown in this subsection, for Liouville-integrable nonlinear vector fields (not necessarily semi-algebraic), our framework can recover the deductive falsification inference while bypassing the \textbf{additional conservatism} due to SOS-based computation.
\end{remark}

%%%%%%%%%%%%%%%%%%%%%%%%%%%%%%%%%%%%%%%%%%%%%%%%%%%%%%%%%%%%%%%%%%%%%%%%%%%%%%%%%%%%%%%%%%%%%%%%%%%%%%%%%%%%%%%%%%%

\section{Effect of Initial Uncertainty}
The inference for probabilistic model validation depends on the initial PDF $\xi_{0}\left(x_{0}\right)$. To account robust inference in the presence of initial PDF uncertainty, the notion of PRVC was introduced in Section 5. However, for many applications, it is desirable to characterize the sensitivity of the gap on the choice of initial PDF. We motivate this issue from two different perspectives.

(i) In predictive modeling applications like systems biology, an important problem is of \emph{model discrimination} \cite{GeorgievKlavins2008,KremlingEtAl2004}, where one looks for an initial PDF that \emph{maximizes} the gap between two models, which seem to exhibit comparable performance. This idea is similar to optimal input design for system identification.

(ii) In general, $_{2}W_{2}\left(t\right) \in \left[0, \sup \parallel y(t) - \widehat{y}(t) \parallel_{2}\right]$, where the supremum is taken over all inter-sample distances between the measured and model-predicted outputs. Thus, $_{2}W_{2}$ is un-normalized and its absolute magnitude may be difficult to interpret when validating a single model against experimental data. Hence, given a set of admissible initial PDFs, it is important to quantify ``\emph{worst-case}" $_{2}W_{2}\left(t\right)$, defined as $\underset{\xi_{0}}{\text{sup}} \: _{2}W_{2}\left(t\right)$, which could be used for normalization.

The main result of this section is that the initial PDF that maximizes Wasserstein distance, depends on the model and true dynamics. In particular, we show that for a linear dynamics pair, the gap is oblivious beyond the first two moments of $\xi_{0}$. We restrict ourselves to scalar dynamics for this analysis.

\subsection{Tools for analysis}
\begin{definition}(\textbf{Quantile function})
Consider the probability space $\left(\Omega_{y},\mathscr{F},\mathbb{P}\right)$ for the output random variable $Y$. Let $y := Y\left(\omega_{y}\right)$, for $\omega_{y} \in \Omega_{y}$. The quantile function $Q_{y}: \Omega_{y} \mapsto \left[0,1\right]$, is defined as the generalized inverse of the CDF for $Y$, i.e.
\begin{eqnarray}
Q_{y}\left(\varsigma\right) := \inf\left(y \in \Omega_{y} : \varsigma \leq \mathbb{P}\left(Y \leq y\right)\right).
\label{QualtileDefn}
\end{eqnarray}
Here $\varsigma \in \left[0,1\right]$ denotes probability mass.
\end{definition}
\begin{proposition}(\textbf{Quantile transport PDE})\cite{SteinbrecherShaw2008}
Consider the scalar SDE $dx\left(t\right) = f\left(x\right) \: dt + g\left(x\right) \: d\beta$, where $\beta$ is the standard Wiener process. Then the \textbf{quantile Fokker-Planck equation} (QFPE), given by
\begin{eqnarray*}
\partial_{t}Q = f\left(Q,t\right) - \displaystyle\frac{1}{2}\partial_{Q}\left(g\left(Q,t\right)\right)^{2} + \displaystyle\frac{1}{2}
\left(g\left(Q,t\right)\right)^{2} \displaystyle\frac{\partial_{\varsigma\varsigma}Q}{\left(\partial_{\varsigma}Q\right)^{2}},
\label{QuantileTransport}
\end{eqnarray*}
describes the transport of quantile function $Q\left(\varsigma,t\right)$ for the process $x(t)$.
\end{proposition}
\begin{proposition}(\textbf{Quantile transformation rule})\cite{Gilchrist2000}
For an algebraic map $y = h\left(x\right)$, we have
\begin{eqnarray}
Q_{y}\left(\varsigma\right) =
\begin{cases} h \circ Q_{x}\left(\varsigma\right) \qquad \text{if} \: h(\cdot) \: \text{is non-decreasing},\\
h \circ Q_{x}\left(1 - \varsigma\right) \; \text{if} \: h(\cdot) \: \text{is non-increasing}.
\end{cases}
\label{QuantileTransformationRule}
\end{eqnarray}
\end{proposition}
Next, we work out some specific results by imposing structural assumptions on the true and model dynamics.

\subsection{Deterministic linear systems}
\begin{figure}[ptbh]
\begin{center}
\includegraphics[width=2.5in]{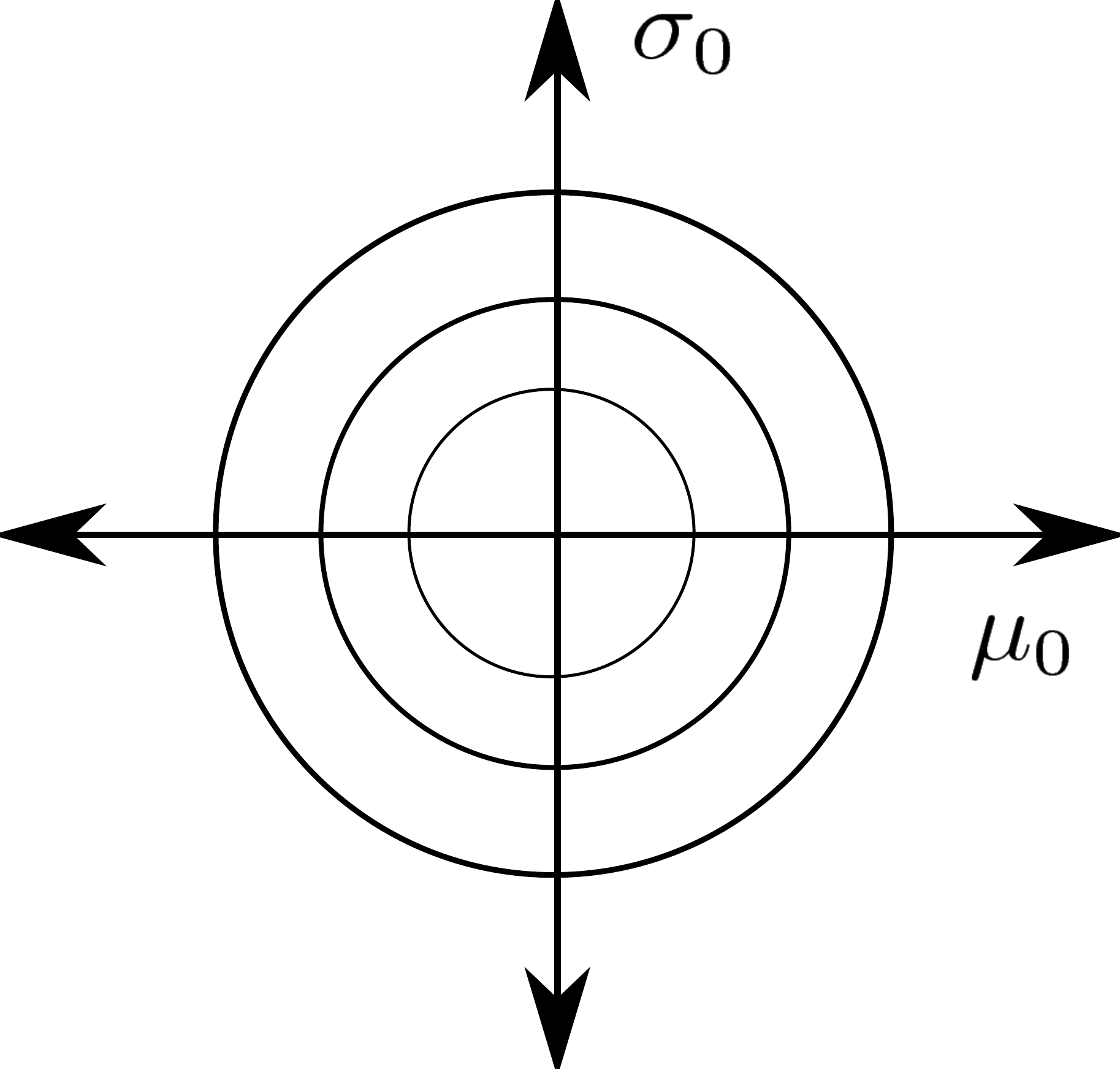}
\end{center}
\caption{\small{The results of Section 4 can be summarized through a graphical algorithm illustrated above.
For scalar linear systems, given a set of admissible initial PDFs over state space, we construct concentric circles centered at origin,
over the two dimensional $\left(\mu_{0},\sigma_{0}\right)$ subspace of the (infinite-dimensional) moment space. From
(\ref{WassClosedFormScalarLinContDet}), $\xi_{0}$ corresponding to the circle with largest radius, maximizes $_{2}W_{2}(t)$, $\forall t>0$. For affine systems, (\ref{AffineWassScalarLinContDet}) implies a similar construction in $\left(\mu_{0},\sigma_{0}\right)$ subspace, with circles
centered at $\left(-\frac{q(t)}{p(t)},0\right)$. The direction of this translation along $\mu_{0}$ axis, depends on parameters
$(a_{i}, b_{i}, c_{i}, d_{i})$, $i=1,2$, of the systems under comparison.}}%
\label{GraphicsInitialPDF}%
\end{figure}

Let the dynamics of the two systems be
\begin{eqnarray}
\dot{x_{i}} = a_{i} x, \quad y_{i} = c_{i}x, \; a_{i} < 0, c_{i} > 0, \qquad i=1,2.
\label{ScalarLinContDet}
\end{eqnarray}
\begin{theorem}
For any initial density $\xi_{0}\left(x_{0}\right)$, the Wasserstein gap between the systems in (\ref{ScalarLinContDet}), is given by
\begin{eqnarray}
_{2}W_{2}\left(t\right) = \sqrt{m_{20}} \; \Big\lvert c_{1}e^{a_{1}t} - c_{2}e^{a_{2}t} \Big\rvert,
\label{WassClosedFormScalarLinContDet}
\end{eqnarray}
where $m_{20} = \mu_{0}^{2} + \sigma_{0}^{2}$, is the second raw moment of $\xi_{0}\left(x_{0}\right)$, while $\mu_{0}$ and $\sigma_{0}$ are its mean and standard deviation, respectively.
\label{ThScalarLinContDet}
\end{theorem}
\begin{pf*}{Proof.}
For (\ref{ScalarLinContDet}), $Q_{y_{i}} = c_{i}Q_{x_{i}}$, and the QFPE reduces to a linear PDE $\partial_{t} Q_{x_{i}} = a_{i}Q_{x_{i}}$, yielding $Q_{x_{i}}\left(\varsigma,t\right) = Q_{0}\left(\varsigma\right) e^{a_{i}t}$, where $Q_{0}$ is the initial quantile function corresponding to $\xi_{0}$. Thus, we have
\begin{eqnarray}
\left(\:_{2}W_{2}\left(t\right)\right)^{2} &=& \displaystyle\int_{0}^{1} \left(Q_{y_{1}}\left(\varsigma,t\right) - Q_{y_{2}}\left(\varsigma,t\right)\right)^{2} \: d\varsigma \nonumber\\
&=& \left(c_{1}e^{a_{1}t} - c_{2}e^{a_{2}t}\right)^{2} \displaystyle\int_{0}^{1} \left(Q_{0}\left(\varsigma\right)\right)^{2} \: d\varsigma.
\label{IntermediateWassScalarLinContDet}
\end{eqnarray}
Since the quantile function maps probability to the sample space, hence $x_{0} = Q_{0}\left(\varsigma\right)$, and $d\varsigma = \xi_{0}\left(x_{0}\right) dx_{0}$. Consequently, we can rewrite (\ref{IntermediateWassScalarLinContDet}) as
\begin{eqnarray*}
\left(\: _{2}W_{2}\left(t\right)\right)^{2} = \left(c_{1}e^{a_{1}t} - c_{2}e^{a_{2}t}\right)^{2} \underset{m_{20}}{\underbrace{\displaystyle\int_{-\infty}^{\infty} x_{0}^{2} \: \xi_{0}\left(x_{0}\right) \: dx_{0}}}.
\end{eqnarray*}
Taking square root to both sides, we obtain the result. It's straightforward to check that $m_{20} = \mu_{0}^{2} + \sigma_{0}^{2}$, relating the central moments with $m_{20}$. \hfill\qed
\end{pf*}
\begin{remark}
(\:\textbf{$_{2}W_{2}$ has limited dependence on $\xi_{0}$})
The above result shows that the Wasserstein gap between scalar linear systems, depends on the initial density up to mean and variance. Any other aspect (skewness, kurtosis etc.) of $\xi_{0}$, even when it's non-Gaussian, has no effect on $_{2}W_{2}\left(t\right)$. The next example demonstrates that our result: ``the initial PDF with maximum second raw moment, maximizes Wasserstein distance" (Fig. \ref{GraphicsInitialPDF}), may be counterintuitive in some situations.
\end{remark}

\begin{figure}[thb]
\begin{center}
\includegraphics[width=3.3in]{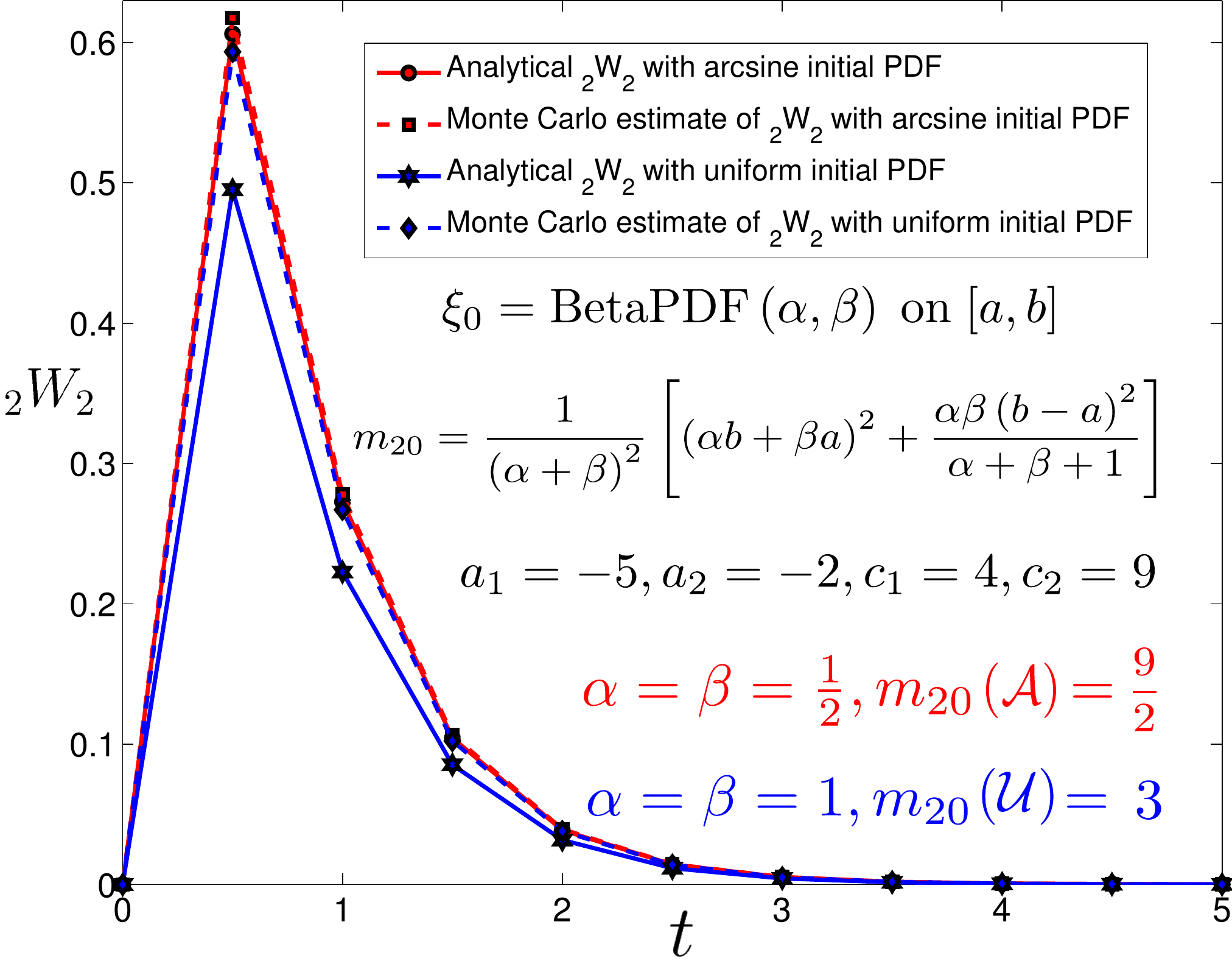}
\end{center}
\caption{\small{Wasserstein time histories between linear system pair (\ref{ScalarLinContDet}) with $\xi_{0}$ as
$\mathcal{A}\left([a,b]\right)$ and $\mathcal{U}\left([a,b]\right)$, respectively. Here $a = -3$, $b = 3$, and
we set sampling interval $\Delta t_{k} = 0.5$. We observe that the Wasserstein gap for
$\xi_{0} = \mathcal{A}\left([a,b]\right)$ remains above the same for $\xi_{0} = \mathcal{U}\left([a,b]\right)$, as
predicted by Theorem \ref{ThScalarLinContDet}. The \emph{solid lines} are direct computation from
(\ref{WassClosedFormScalarLinContDet}), while the \emph{dashed lines} are Monte Carlo estimates of $_{2}W_{2}$ using (\ref{SingleOutputW}).}}%
\label{ExampleScalarInitUncLinearPDF}%
\end{figure}

\begin{example}(\textbf{Uniform initial PDF may not maximize $_{2}W_{2}$})
For (\ref{ScalarLinContDet}), let the set of admissible initial PDFs be $S_{0} := \{\xi_{0} : \text{\emph{supp}}\left(\xi_{0}\right) = \left[a,b\right], \xi_{0}\left(x_{0}\right) = \frac{1}{(b-a)^{\alpha+\beta-1} B\left(\alpha,\beta\right)}(x_{0}-a)^{\alpha-1} (b-x_{0})^{\beta-1}, \alpha>0, \beta>0\}$,
i.e. the set of all scaled beta PDFs supported on $[a,b]$. One can readily compute that $\mu_{0} = \frac{\alpha b + \beta a}{\alpha+\beta}$, and $\sigma_{0}^{2} = \frac{\alpha\beta(b-a)^{2}}{(\alpha+\beta)^{2} (\alpha+\beta+1)}$. For $\alpha=\beta=1$, $\xi_{0} = \mathcal{U}\left([a,b]\right)$, and for $\alpha=\beta=\frac{1}{2}$, $\xi_{0} = \mathcal{A}\left([a,b]\right)$. Thus, we have
\begin{eqnarray}
m_{20}\left(\mathcal{U}[a,b]\right) &=& \frac{1}{3}\left(a^{2} + b^{2} + ab\right), \\
m_{20}\left(\mathcal{A}[a,b]\right) &=& \frac{1}{8}\left(3a^{2} + 3b^{2} + 2ab\right),
\end{eqnarray}
and hence $m_{20}\left(\mathcal{A}[a,b]\right) > m_{20}\left(\mathcal{U}[a,b]\right)$, $\forall \: b > a$.
From Theorem \ref{ThScalarLinContDet}, $_{2}W_{2}(t)$ trajectory for uniform initial PDF, stays below the same for arcsine initial PDF, as shown in Fig. \ref{ExampleScalarInitUncLinearPDF}.
\end{example}

\begin{remark}
(\textbf{Discrete-time linear systems})
Consider the true and model maps $x_{i}^{(k+1)} = a_{i} x_{i}^{(k)}, \, y_{i}^{(k)} = c_{i}x_{i}^{(k)}, \; i=1,2$, where $k\in \mathbb{N}\cup\{0\}$, denotes the discrete time index. From linear recursion, one can obtain a result similar to (\ref{WassClosedFormScalarLinContDet}): $W\left(k\right) = \sqrt{m_{20}} \: \Big\lvert c_{1} a_{1}^{k} - c_{2} a_{2}^{k} \Big\rvert$.
\end{remark}
\begin{remark}
(\textbf{Linear Gaussian systems})
For the linear Gaussian case, one can verify (\ref{WassClosedFormScalarLinContDet}) without resorting to the QFPE. To see this, notice that if $\xi_{0}\left(x_{0}\right)=\mathcal{N}\left(\mu_{0}, \sigma_{0}^{2}\right)$, then the state PDFs evolve as $\xi_{x_{i}}\left(x_{i}, t\right) = \mathcal{N}\left(\mu_{x_{i}}\left(t\right), \sigma_{x_{i}}^{2}\left(t\right)\right)$, where $\mu_{x_{i}}\left(t\right)$ and $\sigma_{x_{i}}^{2}\left(t\right)$ satisfy their respective state and Lyapunov equations, which, in the scalar case, can be solved in closed form. Since $\eta_{y_{i}}\left(y_{i}, t\right) = \mathcal{N}\left(c_{i}\mu_{x_{i}}\left(t\right), c_{i}^{2}\sigma_{x_{i}}^{2}\left(t\right)\right)$, and $_{2}W_{2}$ between two Gaussian PDFs is known \cite{GivensShortt1984} to be $ \sqrt{\left(\mu_{y_{1}} - \mu_{y_{2}}\right)^{2} + \left(\sigma_{y_{1}} - \sigma_{y_{2}}\right)^{2}}$, the result follows.
\end{remark}
\begin{remark}
(\textbf{Affine dynamics}) Instead of (\ref{ScalarLinContDet}), if the dynamics are given by $\dot{x_{i}} = a_{i} x + b_{i}, \; y_{i} = c_{i}x + d_{i}, \; i=1,2$, then by variable substitution, one can derive that $Q_{x_{i}}\left(\varsigma,t\right) = Q_{0}\left(\varsigma\right)e^{a_{i}t} + \displaystyle\frac{b_{i}}{a_{i}}\left(e^{a_{i}t} - 1\right)$. Hence, we get
\begin{eqnarray}
_{2}W_{2}\left(t\right) = \sqrt{\left(p\left(t\right)\right)^{2} m_{20} + 2 p\left(t\right) q\left(t\right) m_{10} + \left(q\left(t\right)\right)^{2}},
\label{AffineWassScalarLinContDet}
\end{eqnarray}
where $m_{10} = \mu_{0}$, $p\left(t\right) := \left(c_{1}e^{a_{1}t} - c_{2}e^{a_{2}t}\right)$, and $q\left(t\right) := \displaystyle\frac{b_{1}c_{1}}{a_{1}}\left(e^{a_{1}t} - 1\right) - \displaystyle\frac{b_{2}c_{2}}{a_{2}}\left(e^{a_{2}t} - 1\right) + \left(d_{1} - d_{2}\right)$.
\end{remark}

\subsection{Stochastic linear systems}
Consider two stochastic dynamical systems with linear drift and constant diffusion coefficients, given by
\begin{eqnarray}
dx_{i} = a_{i} x \: dt + b_{i} \: d\beta, \quad y_{i} = c_{i}x, \qquad i=1,2,
\label{ScalarLinContStoc}
\end{eqnarray}
where $\beta$ is the standard Wiener process.
\begin{theorem}
For any initial density $\xi_{0}\left(x_{0}\right)$, the Wasserstein gap $_{2}W_{2}\left(t\right)$ between the systems in (\ref{ScalarLinContStoc}), is given by
\begin{eqnarray}
_{2}W_{2}\left(t\right) = \sqrt{\left(p\left(t\right)\right)^{2} m_{20} + 2 p\left(t\right) r\left(t\right) s\left(F_{0}\right) + \left(r\left(t\right)\right)^{2}},
\label{WassersteinScalarLinContStoc}
\end{eqnarray}
where $r\left(t\right) := \displaystyle\frac{\lvert b_{1} \rvert c_{1}}{\sqrt{2 a_{1}}}\sqrt{e^{2 a_{1} t} - 1} - \displaystyle\frac{\lvert b_{2} \rvert c_{2}}{\sqrt{2 a_{2}}}\sqrt{e^{2 a_{2} t} - 1}$, and $s\left(F_{0}\right) := \sqrt{2}\:\mathbb{E}\left[x_{0}\:\text{\emph{erf}}^{-1}\left(2F_{0}\left(x_{0}\right) - 1\right)\right]$, $F_{0}$ being the CDF of $x_{0}$.
\end{theorem}
\begin{pf*}{Proof.}
For systems (\ref{ScalarLinContStoc}), quantile functions for the states evolve as (p. 102, \cite{SteinbrecherShaw2008})
\begin{eqnarray}
Q_{x_{i}}\left(\varsigma,t\right) = Q_{0}\left(\varsigma\right) e^{a_{i} t} + \lvert b_{i} \rvert Q_{N}\left(\varsigma\right) \sqrt{\displaystyle\frac{e^{2a_{i}t} - 1}{2a_{i}}},
\end{eqnarray}
where $Q_{N}\left(\varsigma\right) := \sqrt{2}\:\text{erf}^{-1}\left(2\varsigma - 1\right)$, is the standard normal quantile. Thus, the Wasserstein distance becomes
\begin{eqnarray}
\left(\: _{2}W_{2}\left(t\right)\right)^{2} &=& \displaystyle\int_{0}^{1}\left(c_{1}Q_{x_{1}}\left(\varsigma,t\right) - c_{2}Q_{x_{2}}\left(\varsigma,t\right)\right)^{2} \: d\varsigma\nonumber\\
&=&\left(p\left(t\right)\right)^{2} \displaystyle\int_{0}^{1} \left(Q_{0}\left(\varsigma\right)\right)^{2} \: d\varsigma \nonumber\\ &+& 2 p\left(t\right) r\left(t\right) \displaystyle\int_{0}^{1} Q_{0}\left(\varsigma\right) Q_{N}\left(\varsigma\right) \: d\varsigma \nonumber\\ &+& \left(r\left(t\right)\right)^{2} \displaystyle\int_{0}^{1} \left(Q_{N}\left(\varsigma\right)\right)^{2} \: d\varsigma.
\label{IntermediateWassScalarLinContStochastic}
\end{eqnarray}
Notice that the first and third integrals are $m_{20}$ and 1, respectively. Since $\varsigma = F_{0}\left(x_{0}\right)$, the second integral becomes
\begin{eqnarray}
\displaystyle\int_{-\infty}^{\infty} x_{0}\: F_{N}^{-1} \circ F_{0}\left(x_{0}\right) \: \rho_{0}\left(x_{0}\right) dx_{0} \nonumber\\
= \sqrt{2}\:\mathbb{E}\left[x_{0}\:\text{erf}^{-1}\left(2F_{0}\left(x_{0}\right) - 1\right)\right] = s\left(F_{0}\right).
\label{ScalarLinContStocSecondIntegral}
\end{eqnarray}
This completes the proof. \hfill\qed
\end{pf*}
\begin{remark}
(\textbf{Gaussian case})
Consider the special case when $\xi_{0}\left(x_{0}\right) = \mathcal{N}\left(\mu_{0},\sigma_{0}^{2}\right)$. Then $Q_{0}\left(\varsigma\right) = \mu_{0} + \sigma_{0} Q_{N}\left(\varsigma\right)$, and hence the second integral equals $\sigma_{0}$. Thus, if the initial density is normal, then
\begin{eqnarray}
_{2}W_{2} \left(t\right)= \sqrt{\left(p\left(t\right)\right)^{2} m_{20} + 2 p\left(t\right) r\left(t\right) \sigma_{0} + \left(r\left(t\right)\right)^{2}},
\label{WassScalarLinContStocGaussian}
\end{eqnarray}
a function of $\mu_{0}$ and $\sigma_{0}$, which can be verified otherwise by solving the mean and variance propagation equations.
\end{remark}

%%%%%%%%%%%%%%%%%%%%%%%%%%%%%%%%%%%%%%%%%%%%%%%%%%%%%%%%%%%%%%%%%%%%%%%%%%%%%%%%%%%%%%%%%%%%%%%%%%%%%%%%%%%%%%%%%%%

\section{Upper Bounds for $_{2}W_{2}$ for Discrete-time Linear Gaussian Systems}

The objective of this Section is to derive an upper bound of Wasserstein gap for discrete-time linear systems with $\xi_{0}\left(x_{0}\right) = \mathcal{N}\left(0, P_{0}\right)$, in terms of the system matrices and initial covariance. The following result for LTI systems, and its extension for the LTV case have been derived in \cite{HalderBhattacharyaCDC2012}. Here, we only state the LTI result without proof, and then derive a sharper upper bound.
\begin{theorem}
\cite{HalderBhattacharyaCDC2012} Consider two discrete-time stable LTI systems $x_{k+1} = A x_{k}$, and $\widehat{x}_{k+1} = \widehat{A}\widehat{x}_{k}$, $k \in \mathbb{N} \cup \{0\}$. Let the initial PDF $\xi_{0}\left(x_{0}\right) = \mathcal{N}\left(0, P_{0}\right)$. Then, $_{2}W_{2}\left(k\right) \: \leqslant \: \sqrt{2} \left(\text{tr}\left(P_{0}\right)\right)^{1/2} \lvert\lvert \widehat{A}^{-k} \rvert\rvert_{F} \: \Omega_{\text{LTI}}\left(k\right)$, where
{\small{\begin{eqnarray*}
\Omega_{\text{LTI}}\left(k\right)\triangleq\left(\lvert\lvert A^{k} \rvert\rvert_{F}^{2} \lvert\lvert \widehat{A}^{-k} \rvert\rvert_{F}^{2} \left(\text{tr}\left(P_{0}\right)\right)^{2} - \log\left(\prod_{i=1}^{n_{s}}\frac{\vartheta_{i}^{2k}}{\widehat{\vartheta}_{i}^{2k}}\right) - n_{s}\right)^{\frac{1}{2}}
\end{eqnarray*}}}
where the spectrum for $A$ is $\{\vartheta_{i}\}_{i=1}^{n_{s}}$, and for $\widehat{A}$ is $\{\widehat{\vartheta}_{i}\}_{i=1}^{n_{s}}$.
\end{theorem}

\begin{figure}[thb]
\begin{center}
\vspace*{0.2in}
\includegraphics[width=3in]{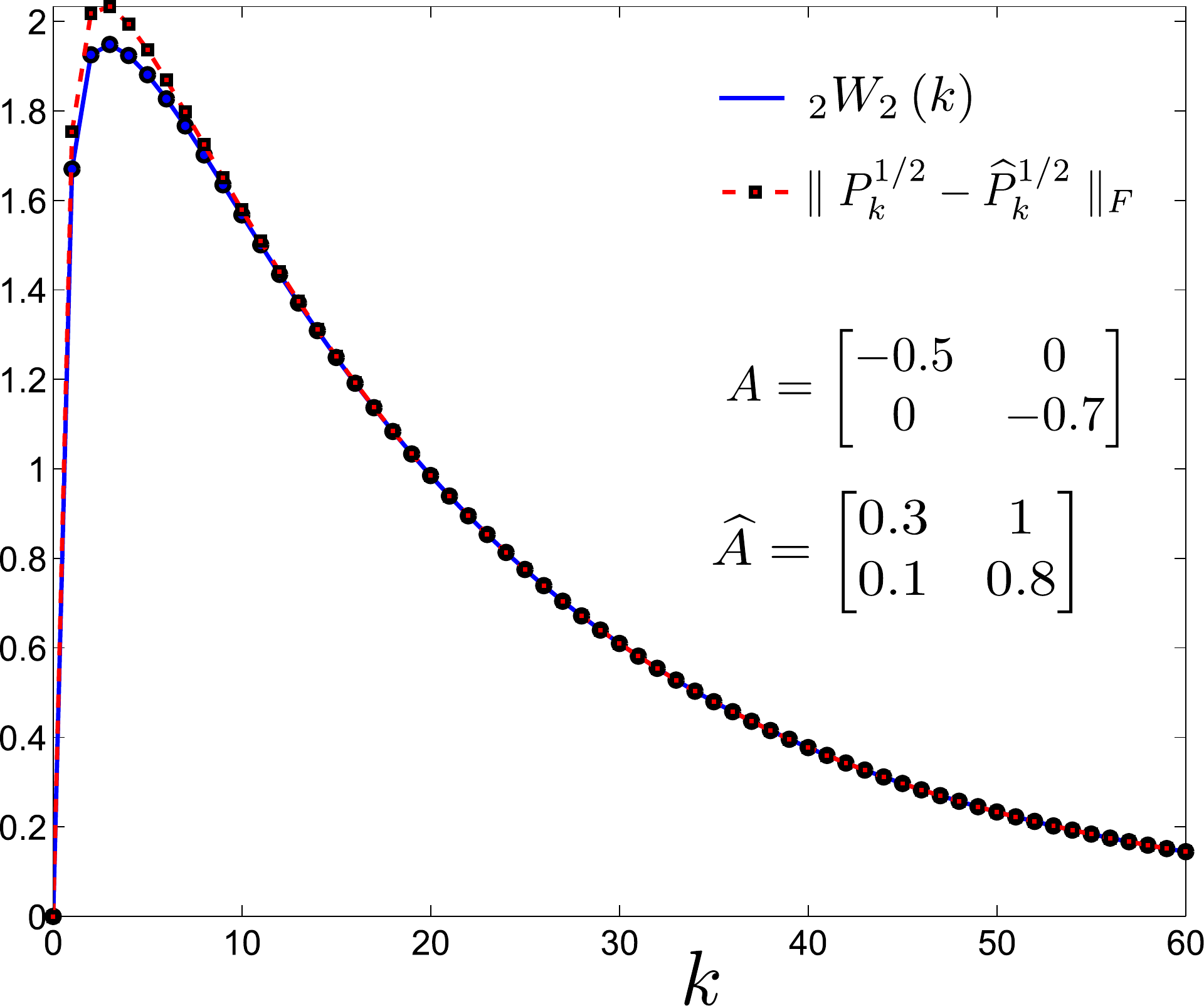}
\end{center}
\caption{\small{Starting with $\mathcal{N}\left(0,P_{0}\right)$, time histories for $_{2}W_{2}(k)$ and its upper bound
(\ref{sharperLTIbound}) for two discrete-time LTI systems, with $A$ and $\widehat{A}$ as shown. Since the systems are stable, both
$_{2}W_{2}(k)$ and $\parallel P_{k}^{1/2} - \widehat{P}_{k}^{1/2} \parallel_{F}$ asymptotically approach zero.}}%
\label{BoundDiscreteTimeLTI}%
\end{figure}

\begin{remark}
(\textbf{A sharper upper bound}) Instead of relating $_{2}W_{2}$ with the spectrum of the LTI systems, one can obtain a sharper bound (see Appendix F for proof):
\begin{eqnarray}
_{2}W_{2}(k) \: \leq \: \parallel P_{k}^{1/2} - \widehat{P}_{k}^{1/2} \parallel_{F},
\label{sharperLTIbound}
\end{eqnarray}
where $P_{k} = A^{k}P_{0}A^{k^{\top}}$, $\widehat{P}_{k} = \widehat{A}^{k}P_{0}\widehat{A}^{k^{\top}}$; the equality holds when $P_{k}$ and $\widehat{P}_{k}$ commute, resulting an interesting Lie bracket condition on system matrices: $\left[A^{k}P_{0}A^{k^{\top}}, \widehat{A}^{k}P_{0}\widehat{A}^{k^{\top}}\right] = 0$. For two Schur-Cohn stable matrices $A$ and $\widehat{A}$, Fig. \ref{BoundDiscreteTimeLTI} illustrates (\ref{sharperLTIbound}) with $P_{0} = \begin{bmatrix} 1 & & 0\\ 0 & & 3\end{bmatrix}$.
\end{remark}

\section{Conclusions}
We have presented a probabilistic model validation framework for nonlinear systems. The notion of soft validation allows us to quantify the degree of mismatch of a proposed model with respect to experimental measurements, thereby guiding for model refinement. A key contribution of this paper is to introduce transport-theoretic Wasserstein distance as a validation metric to measure the difference between distributional shapes over model-predicted and experimentally observed output spaces. The framework presented here applies to any deterministic or stochastic nonlinearity, not necessarily semialgebraic type. In addition to providing computational guarantees for probabilistic inference, we also recover existing nonlinear invalidation results in the literature. Novel results are given for discriminating linear models.

%%%%%%%%%%%%%%%%%%%%%%%%%%%%%%%%%%%%%%%%%%%%%%%%%%%%%%%%%%%%%%%%%%%%%%%%%%%%%%%%%%%%%%%%%%%%%%%%%%%%%%%%%%%%%%%%%%%%%%%%%%

\begin{ack}                               % Place acknowledgements
This research was supported by NSF award \#1016299 with Helen Gill as the program manager. The authors would like to
thank P. Khargonekar at University of Florida, and S. Chakravorty at Texas A\&M University, for insightful discussions. % here.
\end{ack}

%\bibliographystyle{plain}        % Include this if you use bibtex
%\bibliography{autosam}           % and a bib file to produce the
                                 % bibliography (preferred). The
                                 % correct style is generated by
                                 % Elsevier at the time of printing.

\appendix

\section{Computing $_{2}W_{2}\left(\eta_{b}\left(x;\alpha,\beta\right),\eta_{b}\left(x;\beta,\alpha\right)\right)$}
We denote $I_{t}^{-1}\left(\alpha,\beta\right)$ as the inverse of the beta CDF, $I_{x}\left(\alpha,\beta\right) := \displaystyle\frac{B\left(x;\alpha,\beta\right)}{B\left(\alpha,\beta\right)}$ as the regularized incomplete beta function, and $B\left(x;\alpha,\beta\right) := \displaystyle\int_{0}^{x} z^{\alpha-1} \left(1-z\right)^{\beta-1} \: dz$ as the incomplete beta function.
\begin{theorem}
$_{2}W_{2}\left(\eta_{b}\left(x;\alpha,\beta\right),\eta_{b}\left(x;\beta,\alpha\right)\right) = \\ \sqrt{\displaystyle\frac{\alpha\left(\alpha+1\right) + \beta\left(\beta+1\right)}{\left(\alpha+\beta\right)\left(\alpha+\beta+1\right)} - 2 \left(\displaystyle\frac{\beta}{\alpha+\beta} - \mathcal{J}\right)}$,\\
$\mathcal{J} := \displaystyle\frac{1}{\beta+1} \displaystyle\int_{0}^{1} \left(I_{t}^{-1}\left(\alpha,\beta\right)\right)^{1-\alpha} \left(1 - I_{t}^{-1}\left(\alpha,\beta\right)\right)^{1-\beta}$ \\ $\left(I_{t}^{-1}\left(\beta,\alpha\right)\right)^{\beta+1} \: _{2}F_{1}\left(\beta+1, 1-\alpha; \beta+2; I_{t}^{-1}\left(\beta,\alpha\right)\right) \, dt$.
\label{WassersteinDistanceBetweenIsentropicBeta}
\end{theorem}
\begin{pf*}{Proof.}
From (\ref{SingleOutputW}), we have
\begin{eqnarray}
&&_{2}W_{2}^{2}\left(f_{b}\left(x;\alpha,\beta\right),f_{b}\left(x;\beta,\alpha\right)\right) \nonumber\\
&=& \displaystyle\int_{0}^{1} \left(I_{t}^{-1}\left(\alpha,\beta\right) - I_{t}^{-1}\left(\beta,\alpha\right)\right)^{2} \: dt.
\label{BetaWassBasicForm}
\end{eqnarray}
The following identities, stated without proof, will be useful for the evaluation of (\ref{BetaWassBasicForm}).
\begin{property}
\begin{eqnarray*}
\displaystyle\int I_{t}^{-1}\left(a,b\right) \: dt = \displaystyle\frac{1}{\left(a+1\right) B\left(a,b\right)} \left(I_{t}^{-1}\left(a,b\right)\right)^{a+1} \\
_{2}F_{1}\left(a+1, 1-b; a+2; I_{t}^{-1}\left(a,b\right)\right) + \: \text{\emph{constant}}.
\end{eqnarray*}
\label{IndefiniteIntegralInverseRegularizedIncompleteBeta}
\end{property}
\vspace*{-0.2in}
\begin{property}
\begin{eqnarray*}
\displaystyle\int \left(I_{t}^{-1}\left(a,b\right)\right)^{2} \: dt = \displaystyle\frac{1}{\left(a+1\right) B\left(a,b\right)} \left(I_{t}^{-1}\left(a,b\right)\right)^{a+1} \\
\left(\:_{2}F_{1}\left(a+1, 1-b; a+2; I_{t}^{-1}\left(a,b\right)\right) \: - \right.\\
\left. _{2}F_{1}\left(a+1, -b; a+2; I_{t}^{-1}\left(a,b\right)\right)\right) + \: \text{\emph{constant}}.
\end{eqnarray*}
\label{IndefiniteIntegralSquareOfInverseRegularizedIncompleteBeta}
\end{property}
\vspace*{-0.3in}
\begin{property}
\hfill\\
$I_{0}^{-1}\left(a,b\right) = 0$, and $I_{1}^{-1}\left(a,b\right) = 1$.
\label{BoundaryValuesOfInverseRegularizedIncompleteBeta}
\end{property}
\begin{property}
(Gauss Theorem)\hfill\\
$_{2}F_{1}\left(A, B; C; 1\right) = \displaystyle\frac{\Gamma\left(C\right) \Gamma\left(C - A - B\right)}{\Gamma\left(C - A\right) \Gamma\left(C - B\right)}$.
\label{HyperGeomAsRatioOfGammaFunctions}
\end{property}
\begin{property}
\hfill\\
$\displaystyle\frac{d}{dt} I_{t}^{-1}\left(a,b\right) = B\left(a,b\right) \left(I_{t}^{-1}\left(a,b\right)\right)^{1-a} \left(1 - I_{t}^{-1}\left(a,b\right)\right)^{1-b}$.
\label{DerivativeOfInverseRegularizedIncompleteBeta}
\end{property}
Using Properties \ref{IndefiniteIntegralSquareOfInverseRegularizedIncompleteBeta} and \ref{BoundaryValuesOfInverseRegularizedIncompleteBeta}, we get
\begin{eqnarray}
\displaystyle\int_{0}^{1} \left(I_{t}^{-1}\left(\alpha,\beta\right)\right)^{2} \: dt = \displaystyle\frac{1}{\left(\alpha+1\right) B\left(\alpha,\beta\right)} \left[\:_{2}F_{1}\left(\alpha+1, \right.\right.\nonumber\\
\left.\left. 1-\beta;\alpha+2; 1\right) \: - \: _{2}F_{1}\left(\alpha+1, -\beta; \alpha+2; 1\right)\right].\nonumber\\
\label{PauseHere}
\end{eqnarray}
Recalling that $\Gamma\left(k+1\right) = k \Gamma\left(k\right)$, Property \ref{HyperGeomAsRatioOfGammaFunctions} results
\begin{eqnarray}
&& _{2}F_{1}\left(\alpha+1, 1-\beta; \alpha+2; 1\right) = \displaystyle\frac{\Gamma\left(\alpha+2\right) \Gamma\left(\beta\right)}{\Gamma\left(\alpha+\beta+1\right)},\\
&& _{2}F_{1}\left(\alpha+1, -\beta; \alpha+2; 1\right) = \displaystyle\frac{\Gamma\left(\alpha+2\right) \: \beta \Gamma\left(\beta\right)}{\left(\alpha+\beta+1\right) \Gamma\left(\alpha+\beta+1\right)}.\nonumber\\
\label{Simplification}
\end{eqnarray}
Substituting the above expressions in (\ref{PauseHere}), we obtain
\begin{eqnarray}
\displaystyle\int_{0}^{1} \left(I_{t}^{-1}\left(\alpha,\beta\right)\right)^{2} \: dt &=& \displaystyle\frac{\alpha \left(\alpha+1\right)}{\left(\alpha+\beta\right) \left(\alpha+\beta+1\right)},\,\text{similarly}\,\nonumber\\
\displaystyle\int_{0}^{1} \left(I_{t}^{-1}\left(\beta, \alpha\right)\right)^{2} \: dt &=& \displaystyle\frac{\beta \left(\beta+1\right)}{\left(\alpha+\beta\right) \left(\alpha+\beta+1\right)}.
\label{FirstIAndSecondntegral}
\end{eqnarray}
Thus, (\ref{BetaWassBasicForm}) simplifies to
\begin{eqnarray}
&& _{2}W_{2}^{2}\left(\eta_{b}\left(x;\alpha,\beta\right),\eta_{b}\left(x;\beta,\alpha\right)\right) = \displaystyle\frac{\alpha \left(\alpha + 1\right) + \beta \left(\beta + 1\right)}{\left(\alpha + \beta\right) \left(\alpha + \beta + 1\right)} \nonumber\\
&& - 2 \displaystyle\int_{0}^{1} I_{t}^{-1}\left(\alpha, \beta\right) I_{t}^{-1}\left(\beta, \alpha\right) \: dt.
\label{FinalFrontier}
\end{eqnarray}
To evaluate the remaining integral in (\ref{FinalFrontier}), we employ integration-by-parts with $f\left(t\right) := I_{t}^{-1}\left(\alpha, \beta\right)$ as the first function and $g\left(t\right) := I_{t}^{-1}\left(\beta,\alpha\right)$ as the second. Now, we know that $\int_{0}^{1} f\left(t\right) g\left(t\right) dt$ equals
\begin{eqnarray}
 \underbrace{\left[f\left(t\right) \displaystyle\int g\left(t\right) dt\right] \bigg\rvert_{t=0}^{t=1}}_{\mathcal{I}} - \underbrace{\displaystyle\int_{0}^{1} \left(f^{\prime}\left(t\right) \displaystyle\int g\left(t\right) \: dt\right) \: dt}_{\mathcal{J}}.
\label{ByParts}
\end{eqnarray}
From Properties \ref{IndefiniteIntegralInverseRegularizedIncompleteBeta} and \ref{BoundaryValuesOfInverseRegularizedIncompleteBeta}, we get
\begin{eqnarray}
\mathcal{I} &=& \left[\displaystyle\frac{1}{\left(\beta + 1\right) B\left(\alpha, \beta\right)} I_{t}^{-1}\left(\alpha,\beta\right) \left(I_{t}^{-1}\left(\beta, \alpha\right)\right)^{b+1} \right.\nonumber\\
 && \left. _{2}F_{1}\left(b+1, 1-a; b+2; 1\right)\right] \Bigg\rvert_{t=0}^{t=1} \nonumber\\
&=& \displaystyle\frac{ _{2}F_{1}\left(\beta+1, 1-\alpha; \beta+2; 1\right)}{\left(\beta + 1\right) B\left(\alpha, \beta\right)} = \displaystyle\frac{\beta}{\alpha + \beta}.
\label{NiceSimplification}
\end{eqnarray}
Further, Properties (\ref{IndefiniteIntegralInverseRegularizedIncompleteBeta}) and (\ref{DerivativeOfInverseRegularizedIncompleteBeta}) yield
\begin{eqnarray}
&&\mathcal{J} = \displaystyle\frac{1}{\beta+1} \displaystyle\int_{0}^{1} \left(I_{t}^{-1}\left(\alpha,\beta\right)\right)^{1-\alpha} \left(1 - I_{t}^{-1}\left(\alpha,\beta\right)\right)^{1-\beta} \nonumber\\
 && \left(I_{t}^{-1}\left(\beta,\alpha\right)\right)^{\beta+1} \: _{2}F_{1}\left(\beta+1, 1-\alpha; \beta+2; I_{t}^{-1}\left(\beta,\alpha\right)\right) \, dt.\nonumber\\
 \label{CanWeSimplifyThis}
\end{eqnarray}
Combining (\ref{FinalFrontier}), (\ref{ByParts}), (\ref{NiceSimplification}) and (\ref{CanWeSimplifyThis}), the result follows. \hfill\qed
\end{pf*}

\section{On the stationary density of nonlinear systems with multiple stable equilibria}
\begin{proposition}
Consider a nonlinear dynamical system $\dot{x}(t) = f\left(x(t)\right)$, having multiple stable equilibria $\{x_{i}^{\star}\}_{i=1}^{n^{\star}}$. Let us assume that the system does not admit any invariant set other than these stable equilibria. Also, let $\mathcal{R}_{i}$ be the region-of-attraction for the $i$\textsuperscript{th} equilibrium point. If the dynamics evolves from an initial PDF $\xi_{0}$, then its stationary PDF is given by
\begin{eqnarray}
\xi_{\infty}(x) = \displaystyle\sum_{i=1}^{n^{\star}} m_{i}^{\star} \delta\left(x - x_{i}^{\star}\right),
\label{StationaryPDF}
\end{eqnarray}
where $m_{i}^{\star} = \displaystyle\int_{\text{\emph{supp}}\left(\xi_{0}\right)\cap\mathcal{R}_{i}} \xi_{0}\left(x_{0}\right) dx_{0}$.
\end{proposition}
\begin{pf*}{Proof.}
Since $\{x_{i}^{\star}\}_{i=1}^{n^{\star}}$ is the unique set of attractors, it is easy to verify that the stationary PDF is of the form (\ref{StationaryPDF}); however, it remains to determine the weights $m_{i}^{\star}$. We observe that \textbf{either} $\text{supp}\left(\xi_{0}\right) \subseteq \mathcal{R}_{i}$, for some $i = 1, \hdots, n^{\star}$, \textbf{or} $\text{supp}\left(\xi_{0}\right)$ intersects multiple $\mathcal{R}_{i}$.

Now, recall that $\mathcal{R}_{i}\triangleq\{x_{0} : \dot{x}(t) = f\left(x(t)\right), x(0) = x_{0}, \lim_{t\rightarrow\infty}x\left(t\right) = x_{i}^{\star} \}$. Thus, if $\text{supp}\left(\xi_{0}\right) \subseteq \mathcal{R}_{i}$, then $m_{i}^{\star} = \int_{\text{supp}\left(\xi_{0}\right)} dm_{0} = \int_{\text{supp}\left(\xi_{0}\right)} \xi\left(x_{0}\right) dx_{0} = 1$, and consequently, $m_{j}^{\star} = 0$, $\forall j=1,\hdots,n^{\star}$, $j \neq i$, since $\int \xi_{\infty}\left(x\right) dx = 1$. In this case, notice that $\text{supp}\left(\xi_{0}\right) = \text{supp}\left(\xi_{0}\right)\cap\mathcal{R}_{i}$.

On the other hand, if $\text{supp}\left(\xi_{0}\right)$ intersects multiple $\mathcal{R}_{i}$, then only for $x_{0} \in \text{supp}\left(\xi_{0}\right)\cap\mathcal{R}_{i}$, the integral curves of $\dot{x}(t) = f\left(x(t)\right), x(0) = x_{0}$, will satisfy $\lim_{t\rightarrow\infty}x\left(t\right) = x_{i}^{\star}$. In other words, only the set $\text{supp}\left(\xi_{0}\right)\cap\mathcal{R}_{i}$ contributes to $m_{i}^{\star}$, i.e. $m_{i}^{\star} = \int_{\text{supp}\left(\xi_{0}\right)\cap\mathcal{R}_{i}} dm_{0} = \int_{\text{supp}\left(\xi_{0}\right)\cap\mathcal{R}_{i}} \xi\left(x_{0}\right) dx_{0} < 1$.

Combining the above two cases, we conclude $m_{i}^{\star} = \displaystyle\int_{\text{supp}\left(\xi_{0}\right)\cap\mathcal{R}_{i}} \xi_{0}\left(x_{0}\right) dx_{0}$.\hfill \qed
\end{pf*}

\section{Proof for Lemma \ref{RandomVariableInequality}}
Since $X\left(\omega\right) \leqslant Y\left(\omega\right) + Z\left(\omega\right), \: \forall \: \omega \in \Omega$, hence we have $\{\omega: X\left(\omega\right) > \epsilon\} \subseteq \{\omega: Y\left(\omega\right) + Z\left(\omega\right) > \epsilon\} \subseteq \{\omega: Y\left(\omega\right) > \frac{\epsilon}{2}\} \cup \{\omega: Z\left(\omega\right) > \frac{\epsilon}{2}\}$, $\forall \: \omega \in \Omega$. Thus, we get $\mathbb{P}\left(X > \epsilon\right) \: \leqslant \: \mathbb{P}\left(\{Y > \frac{\epsilon}{2}\} \cup \{Z > \frac{\epsilon}{2}\}\right)\: \leqslant \: \mathbb{P}\left(Y > \frac{\epsilon}{2}\right) +\mathbb{P}\left(Z > \frac{\epsilon}{2}\right)$, from Boole-Bonferroni inequality (Appendix C, \cite{MotwaniRaghavan1995}).
\hfill \qed

\vspace*{-0.1in}

\section{Proof for Theorem \ref{WassEstimateUpperBound}}
Since Wasserstein distance is a metric, from triangle inequality
\begin{eqnarray*}
 &&_{2}W_{2}\left(\eta_{m}, \widehat{\eta}_{n}\right) \leqslant \: _{2}W_{2}\left(\eta_{m}, \eta\right) + \: _{2}W_{2}\left(\widehat{\eta}_{n}, \eta\right) \nonumber\\
 &\leqslant& \: _{2}W_{2}\left(\eta_{m}, \eta\right) \: + \: _{2}W_{2}\left(\widehat{\eta}_{n}, \widehat{\eta}\right) \: + \: _{2}W_{2}\left(\eta, \widehat{\eta}\right) \nonumber\\
 &\Rightarrow& \: _{2}W_{2}\left(\eta_{m}, \widehat{\eta}_{n}\right) \: - \: _{2}W_{2}\left(\eta, \widehat{\eta}\right) \: \leqslant \: _{2}W_{2}\left(\eta_{m}, \eta\right) \: + \: _{2}W_{2}\left(\widehat{\eta}_{n}, \widehat{\eta}\right).
 \end{eqnarray*}
Combining the above with Lemma \ref{RandomVariableInequality}, we have
\begin{eqnarray}
\mathbb{P}\left(\bigg\lvert \: _{2}W_{2}\left(\eta_{m}, \widehat{\eta}_{n}\right) \: - \: _{2}W_{2}\left(\eta, \widehat{\eta}\right)\bigg\rvert > \epsilon\right) \: \leqslant \nonumber\\\: \mathbb{P}\left( _{2}W_{2}\left(\eta_{m}, \eta\right) > \displaystyle\frac{\epsilon}{2}\right) \: + \: \mathbb{P}\left( _{2}W_{2}\left(\widehat{\eta}_{n}, \widehat{\eta}\right) > \displaystyle\frac{\epsilon}{2}\right),
\label{AlmostThere}
\end{eqnarray}
where each term in the RHS of (\ref{AlmostThere}) can be separately upper-bounded using Theorem 1 with $\theta \mapsto \displaystyle\frac{\epsilon}{2}$. Hence the result. \hfill\qed

\section{Derivation of stationary PDF (\ref{StationaryDensityContStoc})}    % Each appendix must have a short title.
We re-write the It$\hat{\text{o}}$ SDE (\ref{ContStochastic}) as
\begin{eqnarray}
\begin{Bmatrix}
dx_{1}\\dx_{2}
\end{Bmatrix} =
\begin{Bmatrix}
x_{2}\\- \displaystyle\frac{\partial}{\partial x_{1}}U\left(x_{1}\right) - c x_{2}
\end{Bmatrix} \: dt +
\begin{Bmatrix}
0\\1
\end{Bmatrix} \: dW,
\label{HamiltonianFormContStoch}
\end{eqnarray}
with $U\left(x_{1}\right) := \frac{1}{2}\left(a x_{1}^{2} - b \cos2x_{1}\right)$. An It$\hat{\text{o}}$ SDE with drift nonlinearity of the form (\ref{HamiltonianFormContStoch}),
admits \cite{Klyatskin2005} stationary PDF $\eta_{\infty}\left(x_{1},x_{2}\right) \propto \exp\left(-\frac{c}{Q}
H\left(x_{1},x_{2}\right)\right)$, where the Hamiltonian function $H\left(x_{1},x_{2}\right) := U\left(x_{1}\right) + \frac{1}{2}x_{2}^{2}$.

\section{Proof for $_{2}W_{2}(k) \leq \parallel P_{k}^{1/2} - \widehat{P}_{k}^{1/2} \parallel_{F}$}
It is known (Fact 8.19.21, \cite{Bernstein2009}) that for $0\leqslant p \leqslant 1$, $\text{tr}\left(P_{k}^{p}\widehat{P}_{k}^{p}\right) \leqslant \text{tr}\left(\widehat{P}_{k}^{1/2} P \widehat{P}_{k}^{1/2}\right)^{p}$. Taking $p=\frac{1}{2}$, we get
\begin{eqnarray}
 \text{tr}\left(P_{k}^{\frac{1}{2}}\widehat{P}_{k}^{\frac{1}{2}}\right) \leqslant \text{tr}\left(\widehat{P}_{k}^{\frac{1}{2}} P \widehat{P}_{k}^{\frac{1}{2}}\right)^{\frac{1}{2}}=\text{tr}\left(P_{k}^{\frac{1}{2}} \widehat{P} P_{k}^{\frac{1}{2}}\right)^{\frac{1}{2}},
 \label{TheProdTerm}
 \end{eqnarray}
 where the last equality follows from the symmetry of Wasserstein distance, and can be separately proved by noting that $\text{tr}\left(\sqrt{M M^{\top}}\right) = \text{tr}\left(\sqrt{M^{\top} M}\right)$ for $M = P_{k}^{1/2}\widehat{P}_{k}^{1/2}$.

 Next, recall that square root of a positive definite matrix is unique, and matrix square root commutes with matrix transpose. Thus, we have
 \begin{eqnarray}
 &&\parallel P_{k}^{\frac{1}{2}} - \widehat{P}_{k}^{\frac{1}{2}} \parallel_{F}^{2}
  \:\triangleq\: \text{tr}\left[\left(P_{k}^{\frac{1}{2}} - \widehat{P}_{k}^{\frac{1}{2}}\right)^{\top}\left(P_{k}^{\frac{1}{2}} - \widehat{P}_{k}^{\frac{1}{2}}\right)\right] \nonumber\\
  &=& \text{tr}\left[\left(P_{k}^{\frac{1}{2}}\right)^{\top} P_{k}^{\frac{1}{2}} - \left(P_{k}^{\frac{1}{2}}\right)^{\top} \widehat{P}_{k}^{\frac{1}{2}} - \left(\widehat{P}_{k}^{\frac{1}{2}}\right)^{\top} \widehat{P}_{k}^{\frac{1}{2}} + \left(\widehat{P}_{k}^{\frac{1}{2}}\right)^{\top} \widehat{P}_{k}^{\frac{1}{2}} \right] \nonumber\\
  &=& \text{tr}\left[P_{k}\right] + \text{tr}\left[\widehat{P}_{k}\right] - 2 \: \text{tr}\left[P_{k}^{\frac{1}{2}} \widehat{P}_{k}^{\frac{1}{2}}\right] \nonumber\\
  & \geq & \underbrace{\text{tr}\left[P_{k}\right] + \text{tr}\left[\widehat{P}_{k}\right] - 2 \: \text{tr}\left(P_{k}^{\frac{1}{2}} \widehat{P} P_{k}^{\frac{1}{2}}\right)^{\frac{1}{2}}}_{\left( _{2}W_{2}(k)\right)^{2}} \qquad \text{(using (\ref{TheProdTerm}))}\nonumber
  \label{MakingTheBound}
 \end{eqnarray}
 and hence, $_{2}W_{2}(k) \leq \parallel P_{k}^{1/2} - \widehat{P}_{k}^{1/2} \parallel_{F}$. From (\ref{TheProdTerm}), the equality condition is $P_{k} \widehat{P}_{k} = \widehat{P}_{k} P_{k}$. \hfill \qed

\end{document}